\documentclass[prb,aps,twocolumn,amsmath,amssymb,showpacs]{revtex4}

\usepackage{dcolumn}
\usepackage{bm}
\usepackage{graphicx}
\usepackage{psfrag}

\begin{document}

\title{       Doping dependence of spin and orbital correlations
              in layered manganites }

\author {     Maria Daghofer }
\affiliation{ Institute of Theoretical and Computational Physics,
              Graz University of Technology,
              Petersgasse 16, A-8010 Graz, Austria               \\
              and Max-Planck-Institut f\"ur Festk\"orperforschung,
              Heisenbergstrasse 1, D-70569 Stuttgart, Germany }

\author {     Andrzej M. Ole\'{s} }
\affiliation{ Marian Smoluchowski Institute of Physics, Jagellonian
              University, Reymonta 4, PL-30059 Krak\'ow, Poland  \\
              and Max-Planck-Institut f\"ur Festk\"orperforschung,
              Heisenbergstrasse 1, D-70569 Stuttgart, Germany }

\author {     Danilo R. Neuber }
\author {     Wolfgang von der Linden }
\affiliation{ Institute of Theoretical and Computational Physics,
              Graz University of Technology,
              Petersgasse 16, A-8010 Graz, Austria }

\date{\today}

\begin{abstract}
We investigate the interplay between spin and orbital correlations in
monolayer and bilayer manganites using an effective spin-orbital $t$-$J$
model which treats explicitly the $e_g$ orbital degrees of freedom
coupled to classical $t_{2g}$ spins. Using finite clusters with periodic
boundary conditions, the orbital many-body problem is solved by exact
diagonalization, either by optimizing spin configuration at zero
temperature, or by using classical Monte-Carlo for the spin subsystem at
finite temperature. In undoped two-dimensional clusters, a complementary
behavior of orbital and spin correlations is found --- the ferromagnetic
spin order coexists with alternating orbital order, while the
antiferromagnetic spin order, triggered by $t_{2g}$ spin superexchange,
coexists with  ferro-orbital order. With finite crystal field term, we
introduce a realistic model for La$_{1-x}$Sr$_{1+x}$MnO$_4$, describing
a gradual change from predominantly out-of-plane $3z^2-r^2$ to in-plane
$x^2-y^2$ orbital occupation under increasing doping. The present
electronic model is sufficient to explain the stability of the CE phase
in monolayer manganites at doping $x=0.5$, and also yields the
$C$-type antiferromagnetic phase found in Nd$_{1-x}$Sr$_{1+x}$MnO$_4$
at high doping. Also in bilayer manganites magnetic phases and the
accompanying orbital order change with increasing doping. Here the
model predicts $C$-AF and $G$-AF phases at high doping $x>0.75$,
as found experimentally in La$_{2-2x}$Sr$_{1+2x}$Mn$_2$O$_7$.

(Published in Phys. Rev. B {\bf 73}, 104451, 2006.)
\end{abstract}

\pacs{75.47.Lx, 71.30.+h, 75.10.Lp, 75.75.+a}

\maketitle

\section{Introduction}
\label{sec:intro}

Colossal magnetoresistance (CMR) manganites are characterized by a
complex interplay of charge, spin, orbital and lattice degrees of
freedom. Although manganese oxides have been known for more than 50
years,\cite{Jon50} their properties are still not adequately
understood. After the observation of the CMR effect,\cite{Sch95} the
modeling of complex behavior in this class of compounds has become the
focus of intense research activity in the modern condensed matter
theory.\cite{Dag01,Dag02,Feh04} These recent studies
demonstrate that one has to go beyond the simple double exchange (DE)
model of Zener\cite{Zen51} in order to investigate a complex interplay
between magnetic, orbital, charge and lattice degrees of freedom.

Similar to perovskite manganites, monolayer La$_{1-x}$Sr$_{1+x}$MnO$_4$
and bilayer La$_{2-2x}$Sr$_{1+2x}$Mn$_2$O$_7$ manganites have
interesting and still poorly understood physical properties.
The undoped monolayer LaSrMnO$_4$ compound has the same magnetic
structure as K$_2$NiF$_4$, i.e., it exhibits an $G$-type
antiferromagnetic ($G$-AF) order within the layers.\cite{Mor95,Sen05}
This suggests a different orbital state than that realized in
three-dimensional (3D) LaMnO$_3$ perovskite, where $A$-type
antiferromagnetic ($A$-AF) order, with ferromagnetic (FM) $ab$ planes
and antiferromagnetic (AF) order along $c$ direction, is observed.
Furthermore, unlike in La$_{1-x}$Sr$_{x}$MnO$_3$, in doped monolayer
La$_{1-x}$Sr$_{1+x}$MnO$_4$ compounds no FM metallic phase was observed,
but instead short-range magnetic correlations of various types were
reported,\cite{Ste96,Lar05,Sen05} indicating frustrated magnetic
interactions. This behavior is puzzling and was not explained by the
theory so far.

Also in bilayer La$_{2-2x}$Sr$_{1+2x}$Mn$_2$O$_7$ manganites a
competition between magnetic interactions of different origin was
observed, resulting in rather complex phase diagram,\cite{Med99,Lin00}
which is a challenge for the theoretical models. At higher doping
$x\sim 0.45$ the magnetic order changes from FM to $A$-AF phase.
The observed phase transitions have been ascribed both experimentally
\cite{Koi01} and theoretically\cite{Mae00,Ole03} to the varying crystal
field splitting between $e_g$ orbitals under increasing doping.

In spite of certain similarities between monolayer and bilayer
compounds,\cite{Yam00} the magnetic correlations close to half doping
($x\sim 0.5$) are different. A metallic FM phase is observed in bilayer
manganites up to $x\sim 0.45$,\cite{Per01} while it is absent in
monolayer compounds. While the CE-type AF order is quite pronounced at
$x\simeq 0.5$ in monolayer manganites,\cite{Mur98} the $A$-AF phase is
instead more stable in bilayer systems,\cite{Med99,Lin00} and the
charge order and orbital bi-stripes were also observed for higher
doping $0.55<x<0.6$.\cite{Bea05}

In the present paper, we intend to focus on the role of orbital degrees
of freedom in stabilizing various types of magnetic order observed in
monolayer and bilayer manganites. The behavior of $e_g$ electrons is
dominated by large Coulomb interaction $U$. Therefore, we employ an
effective spin-orbital $t$-$J$ model similar to that derived for the
one-dimensional (1D) chain,\cite{Dag04} and generalize it to the present
situations. Thereby, we implement also the Hund's exchange interaction
$J_H$ which enforces the spin $s=1/2$ of an $e_g$ electron to follow the
$t_{2g}$ spin $S=3/2$ at each site in the ground state. Unlike in the 1D
case, the orbital $e_g$ flavor is not conserved,\cite{Fei05} which
enhances quantum fluctuations. They contribute to intersite correlations
and we show that a close relationship between orbital and spin
correlations nevertheless persists in the ground state and at low
temperature.

The previous theoretical studies revealed a competition between
different types of magnetic order. Among them the most spectacular ones
are phases with coexisting FM and AF bonds:
 (i) $E$-type AF phase in undoped systems,\cite{Dag03,Hot03} and
(ii) the CE phase at half doping ($x=0.5$).
The former one has been experimentally observed for the very strongly
Jahn-Teller (JT) distorted case only, which is at variance with its
theoretical prediction for undistorted compounds. The mechanism of
stability of the CE phase is also still under debate. While it has been
shown that local JT distortions induce the CE-type AF order,
\cite{Cuo02,Bal05} it remains unclear whether it could follow from
electronic interactions alone. It was argued before that this complex
type of magnetic and orbital order (OO) might either originate from
conflicting phases,\cite{Kha99} or could be stabilized by intersite
Coulomb interactions.\cite{Cuo02} Here we address these various
mechanisms proposed before by presenting the evidence obtained by
numerical simulations of finite clusters within a realistic electronic
model including Coulomb interaction. We also show that this model gives
a satisfactory description of magnetic correlations over the
entire doping range.

The paper is organized as follows: In Sec. II we present the effective
$t$-$J$ orbital model in the regime of large $U$ for $e_g$ electrons,
moving in two-dimensional (2D) clusters simulating monolayer manganites,
or in $\sqrt{8}\times\sqrt{8}\times 2$ clusters standing for bilayer
manganites. We also present shortly two numerical methods: the exact
diagonalization with Lanczos algorithm used to solve the orbital model
for fixed spin configurations at zero temperature ($T=0$), and its
combination with Monte Carlo simulations of spin core ($t_{2g}$)
configurations, which leads to a coupled spin-orbital problem at $T>0$.
In Sec. III the model for monolayer manganites is analyzed in different
doping regimes. We report the phase diagrams obtained for undoped and
half doped systems, and relate the obtained magnetic phases to orbital
occupations and intersite orbital correlations. Thereby we highlight
the interrelation between spin and orbital order and their dependence
on increasing doping. The study of bilayer manganites in Sec. IV is
limited by the size of Hilbert space for the smallest relevant
$\sqrt{8}\times\sqrt{8}\times 2$ clusters, so we discuss only undoped
($x=0$), half doped ($x=0.5$) and strongly doped ($x>0.8$) systems.
Finally, in Sec. V we summarize the numerical results and present
general conclusions deduced from the present study for the physical
mechanisms operating in layered manganites.

\section{The model and numerical methods}
\label{sec:model}

\subsection{Orbital $t$-$J$ Model}
\label{sec:otj}

The effective orbital $t$-$J$ model described below follows
from the model of interacting $e_g$ electrons,
\begin{equation}
\label{Heg}
{\cal H}_0=H_t^{(0)}+H_z^{(0)}+H_{\rm int}.
\end{equation}
The form of the kinetic energy $H_t^{(0)}$ depends on the selected basis
of orthogonal orbitals, as discussed in detail by Feiner and Ole\'s.
\cite{Fei05} Here we use the conventional basis which consists of
\begin{equation}
\label{realorbs}
\textstyle{
|z\rangle\equiv \frac{1}{\sqrt{6}}(3z^2-r^2),
\hspace{0.7cm}
|x\rangle\equiv \frac{1}{\sqrt{2}}(x^2-y^2),}
\end{equation}
orbitals. In a 3D (or bilayer) manganite the kinetic energy takes the
form,
\begin{eqnarray}
H_t^{(0)} &=& -\frac{1}{4}t\!\sum_{\langle ij\rangle\parallel ab,\sigma}
  \big[3c_{ix\sigma}^{\dagger}c_{jx\sigma}^{}
       +c_{iz\sigma}^{\dagger}c_{jz\sigma}^{}     \nonumber  \\
  & &\hskip 1.7cm\mp\sqrt{3} (c_{ix\sigma}^{\dagger}c_{jz\sigma}^{}
                            +c_{iz\sigma}^{\dagger}c_{jx\sigma}^{})\big]
                                                  \nonumber  \\
    &-& t\sum_{\langle ij\rangle\parallel c,\sigma}
    c_{iz\sigma}^{\dagger}c_{jz\sigma}^{},
\label{Ht0}
\end{eqnarray}
and the last term is absent for a monolayer. The largest hopping
element $t$ stands for an effective $(dd\sigma)$ element between two
$|z\rangle$ orbitals along the $c$ axis, and originates from two
consecutive $d$-$p$ transitions over oxygen orbital between the
neighboring Mn ions. The same hopping element $t$ couples two equivalent
directional orbitals along either $a$ or $b$ axis, i.e., $3x^2-r^2$ or
$3y^2-r^2$ orbitals.\cite{Fei05}

By considering the structural data, a uniform crystal field splitting
of $e_g$ orbitals is expected both for a monolayer,\cite{Sen05} and
for a bilayer system.\cite{Koi01} Therefore, we introduce the term,
\begin{equation}
H_z^{(0)}=\frac{1}{2}E_z\sum_{i\sigma}(n_{ix\sigma}-n_{iz\sigma}),
\label{Hz0}
\end{equation}
where $n_{i\alpha\sigma}=c_{i\alpha\sigma}^{\dagger}c_{i\alpha\sigma}^{}$
is the electron number operator in $\alpha=x,z$ orbital with spin
$\sigma$ at site $i$. This term describes the crystal field splitting of
$e_g$ orbitals which follows from the geometry of layered manganites and
removes the orbital degeneracy. If $E_z>0$, as in
undoped LaSrMnO$_4$,\cite{Sen05} the $|z\rangle$ orbitals are favored.

The electron interactions between $e_g$ electrons are given by,
\begin{eqnarray}
\label{Hee}
H_{\rm int}&=&
   U\sum_{i,\alpha=x,z}n_{i\alpha  \uparrow}n_{i\alpha\downarrow}
 +\Big(U-\frac{5}{2}J_H\Big)\sum_{i}n_{ix}n_{iz}          \nonumber \\
&+& J_H\sum_{i}
\Big( c^{\dagger}_{ix\uparrow}  c^{\dagger}_{ix\downarrow}
      c^{       }_{iz\downarrow}c^{       }_{iz\uparrow}
     +c^{\dagger}_{iz\uparrow}  c^{\dagger}_{iz\downarrow}
      c^{       }_{ix\downarrow}c^{       }_{ix\uparrow}  \Big)
                                                          \nonumber \\
&-&2J_H\sum_{i}{\vec s}_{ix}\cdot{\vec s}_{iz}
+V\sum_{\langle ij\rangle} n_in_j.
\end{eqnarray}
The spin operators,
${\vec s}_{i\alpha}=\{s_{i\alpha}^+,s_{i\alpha}^-,s_{i\alpha}^z\}$,
are defined by fermion operators in a standard way,
\begin{equation}
s_{i\alpha}^+=c^{\dagger}_{i\alpha\uparrow}c^{}_{i\alpha\downarrow},
\hskip .3cm
s_{i\alpha}^-=c^{\dagger}_{i\alpha\downarrow}c^{}_{i\alpha\uparrow},
\hskip .3cm
s_{i\alpha}^z=\frac{1}{2}(n_{i\alpha\uparrow}-n_{i\alpha\downarrow}).
\label{spins}
\end{equation}
The on-site interactions in $H_{\rm int}$ are rotationally invariant
in the orbital space and are thus described with only two parameters:
\cite{Ole83} the Coulomb element $U$, and a Hund's exchange element
$J_H$. The intersite interactions $\propto V$, where each bond
$\langle ij\rangle$ is included only once, do not depend on the
orbital type and thus involve total electron density operators,
$n_{i}=\sum_{\alpha\sigma}n_{i\alpha\sigma}$.

The Hamiltonian (\ref{Heg}) does not include yet the $t_{2g}$ electrons
which could in principle be described again by a similar multiband
Hamiltonian.\cite{Ole05} However, in reality $t_{2g}$ electrons localize
due to large Coulomb interaction $U\gg t$, so it suffices to consider
local spins $S=3/2$ built by three $t_{2g}$ electron spins of either
Mn$^{3+}$ or Mn$^{4+}$ ion. In both cases each $t_{2g}$ orbital is
singly occupied and virtual intersite hopping processes contribute to
the superexchange $\propto J'$. In this parameter regime, we also find
hopping processes of $e_g$ electrons between two Mn$^{3+}$ ions,
$d_i^4d_j^4\rightleftharpoons d_i^5d_j^3$, that lead to $e_g^2$
configurations at site $i$ (with $e_g^2$ electrons either in one or at
two different orbitals) and generate the $e_g$ part of the superexchange
$\propto J=4t^2/U$. When the system is doped, direct hopping processes
are also possible for Mn$^{3+}$--Mn$^{4+}$ pairs, so one finds an
effective $t$-$J$ model, similar in spirit to the spin $t$-$J$ model
derived almost three decades ago from the Hubbard model.\cite{Cha77}

An important difference to the Hubbard model, however, arises due
to Hund's exchange, $-2J_H{\vec S}_i\cdot{\vec s}_i$ which favors high
spin states at each site. Hund's exchange is close\cite{Zaa90} to the
atomic value of $J_H\sim 0.9$ eV, so is sufficiently larger than the
hopping, estimated for the bilayer\cite{Ole03} to be $t\sim 0.48$ eV,
to assume that spins of $e_g$ electrons are aligned with the $S=3/2$
core spin formed by the $t_{2g}$ electrons. Therefore, core spin
determines a local spin quantization axis at each site $i$, the spins
of $e_g$ electrons can be integrated out,\cite{Kol02} and the
effective orbital $t$-$J$ model takes the form,
\begin{equation}
\label{HtJ}
{\cal H(S)}=H_t+H_J+H_{J'}+H_z+H_V.
\end{equation}
This Hamiltonian may be also obtained by generalizing the effective 1D
model of Ref. \onlinecite{Dag04} to doped layered manganites. It depends
not only on the microscopic parameters introduced above, but also on the
actual configuration ${\cal S}$ of $t_{2g}$ spins on the lattice which
determine the hopping term $H_t$ by the double exchange mechanism.

The first term $H_t$ in Eq. (\ref{HtJ}) stands for the hopping of $e_g$
electrons in the limit of large on-site Coulomb interaction $U\gg t$.
Its form is given below for the monolayer and bilayer system separately.
We emphasize that by performing rigorous projection to the $U\to\infty$
limit\cite{Fei05} the orbital degree of freedom of $e_g$ electrons
survives, but in agreement with the basic idea of the spin $t$-$J$ model
\cite{Cha77} and in contrast to Eq. (\ref{Ht0}), the hopping processes
are now limited to the subspace without (intraorbital and interorbital)
double occupancies. The double occupancies generated in virtual charge
excitations by either $e_g$ or $t_{2g}$ electrons contribute in second
order of the perturbation theory and give the superexchange interactions
$\propto J$ or $\propto J'$, respectively, contained in $H_J$ and
$H_{J'}$ terms.

Similar to LaMnO$_3$,\cite{Fei99} the superexchange in undoped
LaSrMnO$_4$ and in La$_{2-2x}$Sr$_{1+2x}$Mn$_2$O$_7$ is given by a
superposition of several terms. The $e_g$ term $H_J$ favors either FM or
AF spin order on a bond $\langle ij\rangle$, depending on the pair of
occupied $e_g$ orbitals at neighboring sites $i$ and $j$. This makes the
form of the superexchange term $H_J$ depend both on the actual
geometry in layered manganites, and on the used orbital basis. As the
hopping term $H_t$, its form depends on the system and is given below.

In contrast, the superexchange interactions induced by charge
excitations of $t_{2g}$ electrons $H_{J'}$ are identical for planar and
bilayer manganites. They are frequently treated as an effective AF
superexchange between $S=3/2$ core spins, although they couple
{\it de facto\/} two manganese ions in high-spin configurations, and
thus depend on the actual total number of $d$ electrons at both ions.
\cite{Ole02} We have verified, however, that the $t_{2g}$ superexchange
terms derived for these different configurations are of the same order
of magnitude, so in a good approximation one may indeed simulate their
effect by the Heisenberg Hamiltonian with an average exchange constant
$J'>0$ between $S=3/2$ core spins which favors AF spin order; this
interaction is described by the term,
\begin{equation}
\label{HJ'}
H_{J'}=J'\sum_{\langle ij\rangle}
\big({\vec S}_i\cdot{\vec S}_{j}-S^2\big).
\end{equation}
For convenience, we use classical core spins ${\vec S}_i$ of unit length
(compensating their physical value $S=3/2$ by a proper increase of $J'$),
i.e., we replace the scalar products of spin operators on each bond by
their average values. The classically treated ${\vec S}_i$ are
represented by polar angles $\{\vartheta_i,\phi_i\}$
--- then the spin product on a bond $\langle ij\rangle$ is given by:
\begin{equation}
\label{sij}
\langle {\vec S}_i\cdot{\vec S}_{j}\rangle=S^2\big(2|u_{ij}|^2-1\big),
\end{equation}
where the spin orientation enters via
\begin{eqnarray}
\label{uij}
u_{ij}&=&
\cos\Big(\frac{\vartheta_i}{2}\Big)\cos\Big(\frac{\vartheta_j}{2}\Big) +
\sin\Big(\frac{\vartheta_i}{2}\Big)\sin\Big(\frac{\vartheta_j}{2}\Big)
\;e^{i(\phi_j-\phi_i)}                                      \nonumber \\
&=&\cos\Big(\frac{\theta_{ij}}{2}\Big)\;e^{i\chi_{ij}}\;,
\end{eqnarray}
depending on the angle $\theta_{ij}$ between the two involved spins and
on the complex phase $\chi_{ij}$.

The remaining terms in Eq. (\ref{HtJ}), $H_z$ and $H_V$, stand for the
crystal field splitting of two $e_g$ orbitals caused by geometry and for
the nearest neighbor Coulomb interaction. The first term,
\begin{equation}
\label{Hz}
H_z=\frac{1}{2}E_z\sum_i ({\tilde n}_{ix}-{\tilde n}_{iz}),
\end{equation}
involves projected density operators, ${\tilde n}_{ix}$ and
${\tilde n}_{ix}$, that act in the restricted Hilbert space without
(interorbital and intraorbital) double occupancies. The latter term,
\begin{equation}
\label{HV}
H_V=V\sum_{\langle ij\rangle} {\tilde n}_{i}{\tilde n}_{j}.
\end{equation}
is the intersite Coulomb repulsion which survives as the only term
from Eq. (\ref{Hee}). It plays a role at finite doping where it can
induce charge order. Here ${\tilde n}_{i}$ is the total $e_g$ electron
density operator at site $i$ which acts in the restricted Hilbert space:
\begin{equation}
\label{neg}
{\tilde n}_{i}=
{\tilde n}_{ix}+{\tilde n}_{iz}.
\end{equation}
By construction $\langle{\tilde n}_{i}\rangle\le 1$.
We include this term in half doped manganites and investigate in Secs.
\ref{sec:oo} and \ref{sec:phd} whether it helps to stabilize the CE
phase with coexisting charge order.

\subsection{Monolayer manganites}
\label{hammono}

We start with the general form of the hopping term $H_t$ which follows
from Eq. (\ref{Ht0}) for planar (monolayer) manganites in the limit of
$U\gg t$,
\begin{equation}
H_t^{\rm 2D} = -\sum_{\langle ij\rangle\parallel ab}
    t_{i\alpha,j\beta}
    ({\tilde c}_{i\alpha}^{\dagger}{\tilde c}_{j\beta }^{}
    +{\tilde c}_{j\beta }^{\dagger}{\tilde c}_{i\alpha}^{}),
\label{eq:hopping}
\end{equation}
where an operator
${\tilde c}_{i\alpha}^{\dagger}$ creates an electron in $|\alpha\rangle$
state at site $i$ when it is unoccupied by any other ($\alpha$ or
$\beta$) electron, i.e., implements rigorously the restriction of the
Hilbert space to the subspace with no double occupancies. Furthermore,
Hund's exchange between the itinerant $e_g$ electrons to the $t_{2g}$
core spins is large enough to justify the restriction to the subspace
of $e_g$ electrons parallel to the local $t_{2g}$ spin, i.e., we treat
spinless fermions with an orbital flavor $\alpha=x,z$, see Eq.
(\ref{realorbs}). Electrons with antiparallel spins and double
occupancies are treated in second order perturbation theory, see below.

In agreement with double exchange mechanism,\cite{Zen51,Bri99} the
effective hopping strength is modulated by the scalar product of the
core spins at the respective sites --- it is maximal for parallel spins
and vanishes when spins are antiparallel. Accepting the classical
treatment of intersite correlations between core spins, we use
\begin{equation}\label{eq:hopp_u}
 t_{i\alpha,j\beta} = t_{\alpha\beta} u_{ij}\,,
\end{equation}
with $u_{ij}$ given by Eq. (\ref{uij}). The first factor
$t_{\alpha\beta}$ is the orbital-dependent hopping strength. Its form
depends on the used orbital basis, and for a given pair of orbitals
$\{\alpha,\beta\}$ depends on the direction of the bond
$\langle ij\rangle$, as explained below.

While a representation using three directional orbitals $|\zeta\rangle$
along each cubic direction gives the simplest expression for the hopping
of $e_g$ electrons,\cite{Fei05} it is more convenient to consider here
the fixed orthogonal orbital basis given by Eq. (\ref{realorbs}) in a 2D
geometry, for which the orbital-dependent hopping strength
$t_{\alpha\beta}$ in Eq.~(\ref{eq:hopp_u}) is given by:
\begin{equation}\label{eq:orbfactor}
  t^a_{\alpha\beta} = \frac{t}{4}\left(\!\begin{array}{cc}
      3 & -\sqrt{3} \\ -\sqrt{3} & 1
    \end{array} \!\right), \hskip .2cm
  t^b_{\alpha\beta} = \frac{t}{4}\left(\!\begin{array}{cc}
      3 & +\sqrt{3} \\ +\sqrt{3} & 1
    \end{array} \!\right),
\end{equation}
for a bond along the $a$ and $b$ cubic axis, respectively.
The hopping term takes then the form,
\begin{eqnarray}
H_t^{\rm 2D} &=& -\frac{1}{4}t
\sum_{\langle ij\rangle\parallel a,b} u_{ij}
  \big[3{\tilde c}_{ix}^{\dagger}{\tilde c}_{jx}^{}
       +{\tilde c}_{iz}^{\dagger}{\tilde c}_{jz}^{}     \nonumber \\
 && \hskip 1.2cm \mp\sqrt{3}
 ({\tilde c}_{ix}^{\dagger}{\tilde c}_{jz}^{}
 +{\tilde c}_{iz}^{\dagger}{\tilde c}_{jx}^{}) + \mathrm{h.c.}\big].
\label{Ht}
\end{eqnarray}
The correlated fermion operators
${\tilde c}_{iz}^{\dagger}=c_{iz}^{\dagger}(1-n_{ix})$ and
${\tilde c}_{ix}^{\dagger}=c_{ix}^{\dagger}(1-n_{iz})$
act in the restricted Hilbert space and create a $|z\rangle$
($|x\rangle$) electron only when site $i$ is initially empty. The
hopping term (\ref{Ht}) describes therefore spinless fermions with an
orbital flavor in a restricted Hilbert space, in analogy to the original
$t$-$J$ model in spin space,\cite{Cha77} but with an anisotropic hopping
term. We emphasize that the orbital flavor is not conserved along the
hopping process in dimensions higher than one.\cite{Mac99} This has
important consequences for the phase diagram of the orbital Hubbard
model, and destabilizes the OO in doped manganites.\cite{Fei05}

Similar as for the undoped LaMnO$_3$ (see Ref.~\onlinecite{Fei99}),
the $e_g$ superexchange term for undoped LaSrMnO$_4$ depends on the
pair of occupied $e_g$ orbitals at sites $i$ and $j$ for a given bond
$\langle ij\rangle$. If the spins are in the FM configuration, the
superexchange interactions do not vanish but are simply reduced to
purely orbital interactions which favor alternating directional
($3z^2-r^2$-like) and planar ($x^2-y^2$-like) orbitals along every
cubic direction.\cite{vdB99} An effective orbital superexchange model
presented below originates from the complete spin-orbital model
\cite{Fei99} that included the complete multiplet structure of the
excited $d*5$ and $d*4$ states,\cite{Gri71} and focuses on the orbital
dynamics in the presence of spin fluctuations which influence orbital
superexchange interactions.

The superexchange due to $e_{g}$ electron excitations contains spin
scalar products multiplied by orbital interactions on the bonds,
\cite{Fei99} and the full many-body problem would require treating the
coupled spin and orbital dynamics. Here we study only the {\it orbital
correlations\/} and their consequences for the magnetic order by
replacing the scalar products of spin operators on each bond by their
average values (\ref{sij}), as done before in the 1D model,\cite{Dag04}
which is equivalent to decoupling spins and orbitals in a mean-field
approximation. As a result, one is treating the orbital many-body
problem coupled to the classical spins.

The spin operators are replaced by their expectation values following
Eq. (\ref{uij}). In the present case of a monolayer one finds the $e_g$
superexchange term,
\begin{eqnarray}
\label{HJ}
H_J^{\rm 2D}&=& J\sum_{\langle ij \rangle\parallel ab}\Big\{
\frac{1}{5}\Big(2|u_{ij}|^2+3\Big)\big( 2T_{i}^{\zeta}T_{j}^{\zeta}
-\textstyle{\frac{1}{2}{\tilde n}_i{\tilde n}_{j}} \big) \nonumber \\
&-&\frac{9}{10}\big(1-|u_{ij}|^2\big)
                 {\tilde n}_{i\zeta}{\tilde n}_{j\zeta}  \nonumber \\
&-&\big(1-|u_{ij}|^2\big)
\big[{\tilde n}_{i\zeta}(1\!-\!{\tilde n}_{j})
+(1\!-\!{\tilde n}_i){\tilde n}_{j\zeta}\big]\Big\},
\end{eqnarray}
where number operator ${\tilde n}_{i\zeta}$ refers in each case to the
directional orbital $|\zeta\rangle$ along a given bond
$\langle ij \rangle$, and
\begin{equation}
\label{Tzeta}
T_{i}^{\zeta}=-\textstyle{\frac{1}{2}}\big(T_{i}^z
                                \mp\sqrt{3}T_{i}^x\big),
\end{equation}
depend on the bond direction, with the sign $-$ ($+$) in Eq.
(\ref{Tzeta}) corresponding to $a$ ($b$) axis. The operators are
defined by orbital $T=1/2$ pseudospin operators,
\begin{eqnarray}
\label{Tz}
T_{i}^{z}&=&\textstyle{\frac{1}{2}}\sigma^z_i
=\textstyle{\frac{1}{2}}\big({\tilde n}_{ix}-{\tilde n}_{iz}\big),   \\
\label{Tx}
T_{i}^{x}&=&\textstyle{\frac{1}{2}}\sigma^x
=\textstyle{\frac{1}{2}}
\big({\tilde c}^\dagger_{ix}{\tilde c}^{\phantom{\dagger}}_{iz}
    +{\tilde c}^\dagger_{iz}{\tilde c}^{\phantom{\dagger}}_{ix}\big),
\end{eqnarray}
with two eigenstates of $T_{i}^{z}$, see Eq. (\ref{realorbs}).
The superexchange constant $J=t^2/\varepsilon(^6\!A_1)$ is determined by
the lowest high-spin excitation energy $\varepsilon(^6\!A_1)$ at the
Mn$^{2+}$ ion.\cite{Fei99}

\subsection{Bilayer manganites}
\label{sec: hambila}

The hopping term for the bilayer manganites can be written again in
terms of directional orbital along the bonds, by extending Eq.
(\ref{Ht}). In fact, the interlayer hopping term along $c$-axis is then
diagonal in $\{|x\rangle,|z\rangle\}$ basis,
\begin{equation}
\label{eq:orbfactor_z}
  t^c_{\alpha\beta} = t\left(\begin{array}{cc}
      0 & 0 \\ 0 & 1
    \end{array} \right),\
\end{equation}
and one finds,
\begin{eqnarray}
  H_t^{\rm BL} &=& -\frac{1}{4}t\sum_{\langle ij\rangle\parallel a,b}
  u_{ij}
  \Big[3{\tilde c}_{ix}^{\dagger}{\tilde c}_{jx}^{}
  +{\tilde c}_{iz}^{\dagger}{\tilde c}_{jz}^{}           \nonumber \\
  && \hskip 1.7cm \mp\sqrt{3}
  ({\tilde c}_{ix}^{\dagger}{\tilde c}_{jz}^{}
  +{\tilde c}_{iz}^{\dagger}{\tilde c}_{jx}^{})+ \mathrm{h.c.}\Big]
  \nonumber \\
  &&-t\sum_{\langle ij\rangle\parallel c} u_{ij}\Big(
  {\tilde c}_{iz}^{\dagger}{\tilde c}_{jz}^{} + \mathrm{h.c.}\Big).
  \label{Htbila}
\end{eqnarray}
As for a single plane, the hopping $H_t$ describes spinless fermions
with an orbital flavor in a restricted Hilbert space. In each plane the
orbital flavor is not conserved along the hopping process.\cite{Fei05}

We apply again the same approach described above to the superexchange
interaction; it leads in the present case of a bilayer system to the
following expression for the $e_g$ superexchange,
\begin{eqnarray}
\label{HJbila}
H_J^{\rm BL}&=& J\sum_{\langle ij \rangle\parallel ab}\Big\{
  \frac{1}{5}\Big(2|u_{ij}|^2+3\Big)\big( 2T_{i}^{\zeta}T_{j}^{\zeta}
 -\textstyle{\frac{1}{2}{\tilde n}_i{\tilde n}_{j}} \big)  \nonumber \\
&-&\frac{9}{10}\big(1-|u_{ij}|^2\big)
 {\tilde n}_{i\zeta}{\tilde n}_{j\zeta}                    \nonumber \\
&-&\big(1-|u_{ij}|^2\big)
 \big[{\tilde n}_{i\zeta}(1\!-\!{\tilde n}_{j})
 +(1\!-\!{\tilde n}_i){\tilde n}_{j\zeta}\big]\Big\}       \nonumber \\
&+& J\sum_{\langle ij \rangle\parallel c}\Big\{
 \frac{1}{5}\Big(2|u_{ij}|^2+3\Big)\big( 2T_{i}^{z}T_{j}^{z}
 -\textstyle{\frac{1}{2}{\tilde n}_i{\tilde n}_{j}} \big)  \nonumber \\
&-&\frac{9}{10}\big(1-|u_{ij}|^2\big)
 {\tilde n}_{iz}{\tilde n}_{jz}                            \nonumber \\
&-&\big(1-|u_{ij}|^2\big)
\big[{\tilde n}_{iz}(1\!-\!{\tilde n}_{j})
+(1\!-\!{\tilde n}_i){\tilde n}_{jz}\big]\Big\}.
\end{eqnarray}
Compared to Sec. \ref{hammono}, it is extended by the interlayer
coupling terms, with the charge and orbital operators given by
${\tilde n}_{iz}$ and $T_i^z$ along $c$ axis [the remaining terms for
the bonds in $ab$ planes are the same as in Eq. (\ref{HJ})].
For simplicity, we neglect here the difference between the intralayer
and interlayer hopping elements, although for a detailed comparison with
experiment it might be necessary to include different bond lengths for
each cubic direction.

\subsection{Density distribution and intersite correlations}

Below we present the physical quantities which are used to characterize
electron density distribution and intersite spin and orbital
correlations in the ground states and at low temperature. The density
distribution in both monolayer and bilayer systems at doping $x$ is
described by average electron densities in orbital $\alpha$,
\begin{equation}
n_{\alpha}=\frac{1}{N}\sum_i\langle {\tilde n}_{i\alpha}\rangle=
  \frac{1}{N}\sum_i\langle{\tilde c}^\dagger_{i\alpha}
              {\tilde c}^{\phantom{\dagger}}_{i\alpha}\rangle,
\label{nalpha}
\end{equation}
with $N$ being the number of lattice sites.

\begin{figure}[t!]
\includegraphics[width=8cm]{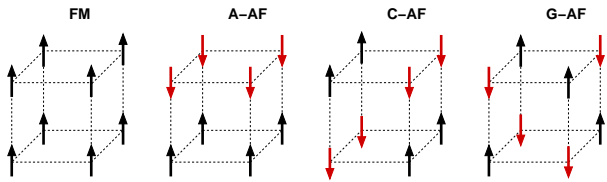}
\caption{(Color online)
Schematic spin structures in the magnetic phases of doped manganites.
The number of FM bonds decreases gradually from FM to $G$-AF phase,
through $A$-AF and $C$-AF phase, with eight and four FM bonds in a cube.
For an $ab$ monolayer, FM and $A$-AF phases are equivalent.
}
\label{fig:acg}
\end{figure}

Depending on the actual parameters, we encountered several different
types of magnetic order, both in monolayer and in bilayer model.
The FM and AF spin interactions compete with each other, so one
of them is selected by a particular type of orbital correlations.
The possible generic types of magnetic order, all observed in the
manganites,\cite{Dag01,Dag02,Feh04} are shown schematically in Fig.
\ref{fig:acg}. We have investigated intersite correlations
at short distance ${\vec r}$ by:
(i) intersite spin correlations, given by the scalar product
    (\ref{sij}) of the classical core spins,
\begin{equation}
{\cal S}(\vec r)=\frac{1}{N}\sum_{i}
\langle {\vec S}_{i}\cdot{\vec S}_{i+\vec r}\rangle\;,
\label{ss}
\end{equation}
and (ii) intersite orbital correlations,
\begin{equation}
{\cal T}_{\theta}(\vec r)=\frac{1}{N}\sum_i
                   \langle T_{i}(\theta)
                           T_{i+\vec r}(\theta)\rangle\;.
\label{tco}
\end{equation}
The orbital operators are defined using a particular orbital basis,
\begin{equation}
T_{i}(\theta)=T_i^z\,\cos\theta+T_i^x\,\sin\theta,
\label{oo}
\end{equation}
and the pseudospin operators $\{T_i^z,T_i^x\}$ at site $i$ are given
by Eqs.~(\ref{Tz}) and (\ref{Tx}). For instance, $\theta=0$ corresponds
to $T_i^zT_{i+m}^z$, $\theta=\pi/2$ --- to $T_i^xT_{i+m}^x$, and
$\theta=2\pi/3$ --- to $T_i^{\zeta}T_{i+{\vec r}}^{\zeta}$, with
$\zeta$ standing for the directional orbital along $a$ axis. The orbital
correlations expected in undoped manganites are of AO type on two
sublattices, which suggests that the orbital correlations
${\cal T}_{\theta}(\vec r)$ defined as in Eq. (\ref{tco}) are
predominantly negative for nearest neighbors. These correlations were
investigated along the (10) and (11) directions in 2D clusters. In the
bilayers we will distinguish between two different cases for both above
directions, and consider: either
 (i) both sites within the same $ab$ layer, or
(ii) two sites which belong to different layers.

Spin correlations in a doped system were uniquely determined by core
spin correlations. When $e_g$ holes are present,
it is also of interest to investigate spin correlations near the hole,
so we show in some cases the correlations between the two nearest
neighbor spins separated by a hole,
\begin{equation}
\label{sns}
{\cal R}=\frac{1}{N}\sum_{i}\langle
      {\vec S}_{i-\vec e}(1-{\tilde n}_i){\vec S}_{i+\vec e}\rangle\;,
\end{equation}
where $(1-\tilde n_i)$ stands for the hole density at the central site
$i$, and two spins ${\vec S}_{i \pm\vec e}$ occupy two adjacent sites
in either $a$ or $b$ direction.

\subsection{Numerical methods}
\label{sec:num}

We employed two different but related numerical methods to investigate
finite clusters described by the orbital $t$-$J$ model:
 (i) exact diagonalization at zero temperature ($T=0$), and
(ii) its combination with Markov chain Monte Carlo (MC) at finite
     temperature $T>0$.
The ground state at $T=0$ for representative electron fillings was
determined by solving the orbital $t$-$J$ model for several possible
types of spin order, using clusters with periodic boundary
conditions: $\sqrt{8}\times\sqrt{8}$ and $4\times 4$ clusters for
monolayer manganites, and $\sqrt{8}\times\sqrt{8}\times 2$ clusters for
bilayer ones. Here, the spin configurations were fixed, corresponding
to different possible magnetic phases: FM, AF, $C$-AF, $E$-AF, and CE
phase. To solve the orbital problem specified by the selected core spin
configuration, we employed the Lanczos algorithm, which is suitable for
very large matrices treated here and guarantees fast convergence to the
ground state in each case. We then determined the global ground state
by comparing the ground-state energies obtained for different magnetic
phases.

At finite temperatures, we investigated the effective orbital $t$-$J$
model by making use of a combination of Markov chain Monte Carlo
algorithm for the core spins\cite{Dag04} with Lanczos diagonalization.
For each classical core spin configuration occurring in the MC runs, we
defined the actual values of classical variables $\{u_{ij}\}$, and next
employed Lanczos diagonalization to solve the many-body problem posed by
the orbital model. In each case we obtained the free energy for that
core spin configuration from the few lowest eigenstates, which was next
used to decide acceptance in the MC runs.
We measured the Boltzmann weight of these lowest eigenstates and thereby
made sure that only states with negligible weight were discarded. We
emphasize that a complete many-body problem in the orbital subspace was
solved, so our method differs from the standard algorithm used for
noninteracting electrons which employs free-fermion formulae, as e.g.
in Ref. \onlinecite{Dag98}.

In the MC updates, the angle of spin rotation was optimized to keep
acceptance high enough. If acceptance was very good, several rotations
were performed in each update. From time to time, a complete spin flip
$\mathbf S_i\to -\mathbf S_i$ was proposed. Autocorrelation analysis
was employed to obtain reliable error estimates and several hundred
effectively uncorrelated samples were considered, taking particular
care of burn-in and thermalization processes. Finally, wherever
autocorrelations were observed to be particularly long, e.g. in
symmetry-broken states like the CE phase, the method of parallel
tempering\cite{Huk95} was employed.

\section{ Numerical results for monolayer manganites }
\label{sec:res}

\subsection{Undoped 2D clusters }
\label{sec:undo}

\begin{figure}[t!]
\includegraphics[width=8.2cm]{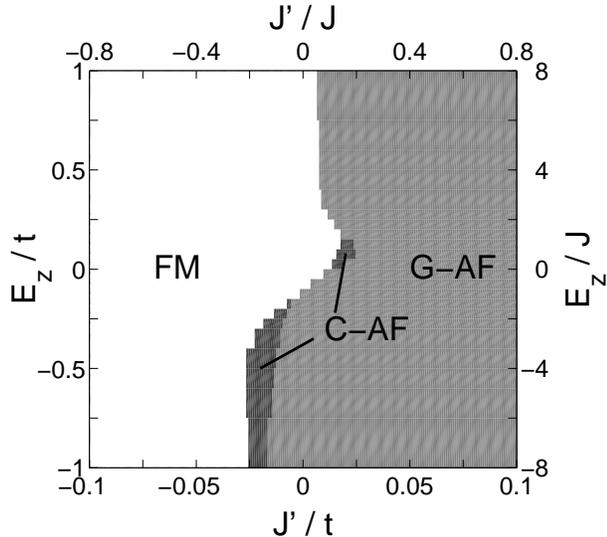}
\caption{
Phase diagram for the undoped layered manganites ($x=0$), with the
stability regions of FM, $G$-AF, and $C$-AF phases (see Fig.
\ref{fig:acg}) in $(J',E_z)$ plane, as obtained at $T=0$ with a
$4\times 4$ cluster.
The core spin superexchange $J'$ and the crystal field splitting $E_z$
are given both in units of $J$, and in units of $t$ for $J=0.125t$.
}
\label{fig:phd1}
\end{figure}

In undoped manganites (at $x=0$), the hopping is entirely blocked by
large Coulomb interaction $U$, and the nearest-neighbor Coulomb
interaction $V$ (\ref{HV}) is irrelevant as the electron density is
$n=1$ at each site. Therefore, the Hamiltonian for a monolayer Eq.
(\ref{HtJ}) reduces to the superexchange terms given by Eqs. (\ref{HJ'})
and (\ref{HJ}), accompanied by the crystal-field splitting of $e_g$
orbitals, given by Eq. (\ref{Hz}). Therefore, the ground state for
finite $J$ is determined by only two parameters: $J'/J$ and $E_z/J$.
The phase diagram of the spin-orbital model (\ref{HtJ}) in $(J',E_z)$
plane obtained by exact diagonalization of a $4\times 4$ cluster at
$T=0$ is shown in Fig. \ref{fig:phd1}.

A particularly simple result is obtained at $J'=E_z=0$, where the FM
phase has the lowest energy. As $u_{ij}\equiv 1$ in a ferromagnet, only
the first term in the $e_g$ superexchange (\ref{HJ}) survives and
stabilizes the alternating orbital (AO) order. Of course, this phase is
also stable in an unphysical regime\cite{notefm} of negative $J'<0$ in
a broad range of crystal field splitting, see Fig. \ref{fig:phd1}. This
coexistence of the AO order with the FM spin correlations is generic
--- it confirms the trend observed before in the 1D model\cite{Dag04}
and on ladders,\cite{Neu06}
and agrees with the Goodenough-Kanamori rules.\cite{Goode}
When robust AO state develops at the orbital degeneracy ($E_z=0$),
the FM phase extends to a broader range of $J'\gtrsim 0$ than at
$|E_z|>0$.

As the FM and AF terms in $e_g$ superexchange (\ref{HJ}) compete with
each other, the $G$-AF phase has only slightly higher energy than the
FM one at $E_z=0$, so it can be rather easily stabilized by finite AF
$t_{2g}$ superexchange $J'>0$.
One finds indeed a transition to the AF order at $J'\gtrsim 0.0091t$
(we use below $t=1$ as an overall energy unit). Note, however, that
near orbital degeneracy ($E_z\sim 0$) robust AO state develops as both
orbitals can participate in equal amount, and therefore the FM phase
extends here to higher values of $J'$ than at large orbital splitting
$|E_z|$. Therefore, the crystal field energy may easily tip the energy
balance and stabilize a different type of magnetic order, although
$E_z$ controls primarily the charge distribution. Indeed, at large
$|E_z|$ only one of $e_g$ orbitals is selected, AO order is hindered,
so the AF phase is favored and occurs already for smaller $J'$ than at
$E_z=0$, see Fig.~\ref{fig:phd1}. For $E_z<0$, $|x\rangle$ orbitals are
occupied and one finds large AF $e_g$ superexchange --- therefore in
this case the AF phase is favored even for $J'=0$. In contrast, when
$|z\rangle$ orbitals are favored for $E_z>0$, the $e_g$ superexchange is
weaker by a
factor of $(t_{xx}/t_{zz})^2=9$, and the $G$-AF phase can be stabilized
only by $J'\simeq 0.01t$. Moreover, one finds that the competition
between the FM and AF order may induce the $C$-AF phase in the crossover
regime between the FM and AF phase. We note that the $C$-AF phase is
expected to be further stabilized by finite cooperative JT distortions.

\begin{figure}[t!]
\includegraphics[width=7.7cm]{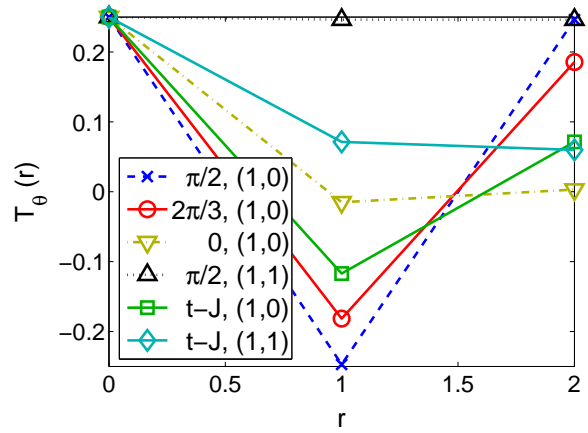}
\caption{(Color online)
Orbital correlations ${\cal T}_{\theta}(\vec r)$ (\ref{tco}) in
$(01)$ direction in the FM phase, as obtained for the undoped
$4\times 4$ cluster using $T_i^z$ operators defined for different
bases of orthogonal orbitals:
$\theta=0$ --- $\{|x\rangle,|z\rangle\}$;
$\theta=\pi/2$ --- $\{(|x\rangle\pm|z\rangle)/\sqrt{2}\}$;
$\theta=2\pi/3$ --- $\{|\zeta\rangle, |\xi\rangle\}$.
The latter choice involves directional orbitals within the $ab$ plane.
Orbital correlations shown in $(11)$ direction for $\theta=\pi/2$
demonstrate an almost perfect AO order. Parameters: $E_z=J'=0$.
Spin correlations obtained for the spin model (i.e. the Heisenberg model
for $s=1/2$) on a $4\times 4$ cluster at $n=1$ are shown for comparison
by squares and diamonds.
}
\label{fig:oo_theta_FM}
\end{figure}

In order to get more insight into the phases of Fig.~\ref{fig:phd1},
we investigated orbital correlations between first and second
neighbors. Let us first look at the ferromagnetic phase at $E_z=0$:
Since there is no hopping at $x=0$ and since all bond are FM
(i.e. $u_{ij}\equiv 1$), the Hamiltonian comprises only the first term
of Eq.~\ref{HJ}. While it looks similar to a spin $t$-$J$ model (or in
this undoped case the Heisenberg model for $s=1/2$), the important
difference is that the orbital model is not isotropic. As a consequence,
the orbital correlations (\ref{tco}) differ markedly from those familiar
from the isotropic spin $t$-$J$ model, and depend on the considered
orbitals basis. They are shown in Fig.~\ref{fig:oo_theta_FM} for various
orbitals parameterized by angle $\theta$ [see Eq. (\ref{oo})], and one
finds that almost no correlations are found when angle $\theta=0$ is
selected, i.e., for ${\cal T}_{ij}(0)=\langle T_{i}^{z}T_{j}^{z}\rangle$.
In contrast, orbital correlations designed to detect the AO order, with
angle $\theta\sim \pi/2$, are quite distinct, as for instance for
$\theta=2\pi/3$, when correlations between two directional
$3x^2-r^2/3y^2-r^2$ orbitals within the plane are measured by
${\cal T}_{ij}(2\pi/3)=T_{i}^{\zeta}T_{j}^{\zeta}$. As this correlation
function is negative, one finds the OO close to staggered directional
orbitals on two sublattices, with $\theta_{i\in A}=\theta$ and
$\theta_{j\in B}=-\theta$, i.e.,
\begin{eqnarray}
\label{ood9}
|\theta\rangle_{i\in A}&=&\cos\Big(\frac{\theta}{2}\Big)|z\rangle_i
                    +\sin\Big(\frac{\theta}{2}\Big)|x\rangle_i,     \\
|\theta\rangle_{j\in B}&=&\cos\Big(\frac{\theta}{2}\Big)|z\rangle_j
                    -\sin\Big(\frac{\theta}{2}\Big)|x\rangle_j.
\end{eqnarray}
Remarkably, the orbital correlations defined by the directional orbitals
on two sublattices $(-0.1816)$ are very similar to those obtained with
${\cal T}_{ij}(2\pi/3)$ (-0.1892), but both types of OO do not minimize
the negative orbital superexchange. A common large negative contribution
to the above values comes from $\langle T_i^xT_j^x\rangle=-0.1854$
correlation --- it decreases further when the angle $\theta$ is varied
towards $\theta=\pi/2$, where the orbital correlations reach a minimal
value, ${\cal T}_{ij}(\pi/2)=\langle T_{i}^{x}T_{j}^{x}\rangle=-0.2472$.
Thus, in agreement with earlier findings,\cite{vdB99,Neu06} the quantum
correction to the classical value $-0.25$ is very small indeed due to
the gap which opens in orbital excitations in the present 2D case. This
means that robust AO order with orbitals of the form
$(|x\rangle\pm|z\rangle)/\sqrt{2}$ is realized in an undoped monolayer
without crystal field splitting, similar to the OO in the 3D model,
\cite{Fei99} if this monolayer has FM spin order.

\begin{figure}[t!]
\includegraphics[width=7.7cm]{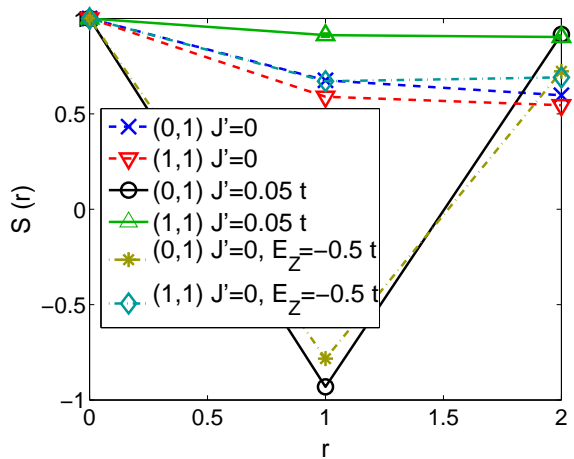}
\caption{(Color online)
Spin correlations ${\cal S}(\vec r)$ (\ref{ss}) in the undoped
monolayer, as obtained in MC simulations with a $4\times 4$ cluster for
$J'=0$ and $E_z=0$ (FM correlations), and for $J'=0$, $E_z=-0.5t$,
as well as for $J'=0.05t$, $E_z=0$ (both sets give AF correlations).
Error bars are smaller than symbol sizes.
Parameter: $J=0.125t$, $\beta t=100$.}
\label{fig:ss_MC_x0_2d}
\end{figure}

For comparison, we also included in Fig.~\ref{fig:oo_theta_FM} the spin
correlation (\ref{ss}) of the spin $t$-$J$-model --- it is isotropic
due to SU(2) symmetry, and quantum fluctuations keep the value of
${\cal S}(1)$ for (10) direction well above the classical value $-1/4$.
In contrast, for the orbital
model the optimal correlations found for $\theta=\pi/2$ almost attain
the classical value and thus show almost perfect OO. An additional
advantage of this robust OO for the present calculations is that the
ground-state energy $E_0$ hardly depends on the cluster size --- one
finds for the FM phase ($u_{ij}=1$) at $E_z=0$:
$E_0=-0.21970t$ for $\sqrt{8}\times\sqrt{8}$,
    $-0.21968t$ for $\sqrt{10}\times\sqrt{10}$,
and $-0.21967t$ for $4\times4$ cluster, i.e., finite-size effects are
negligible.
When the AF order is considered instead ($u_{ij}=0$), the second term in
the orbital superexchange Eq. (\ref{HJ}) dominates and thus almost only
in-plane $|x\rangle$ orbitals are occupied. Finite size effects are here
again small, with:
$E_0=-0.18335t$ for $\sqrt{8}\times\sqrt{8}$,
    $-0.18334t$ for $\sqrt{10}\times\sqrt{10}$, and
    $-0.18333t$ for $4\times 4$ cluster.

In the present model one does not find the $E$-AF phase, which has been
experimentally observed for the strongly JT distorted HoMnO$_3$.
\cite{Kim03} We attribute this to the fact that the present model does
not include phonons and thus rather describes manganites with little or
no JT distortion. The $E$-AF phase was found before in MC simulations
using a model neglecting on-site Coulomb repulsion in the regime of
large $J'$ and small electron-phonon coupling $\lambda$, where it was
stabilized by the kinetic energy.\cite{Hot03,Sal06} It could argued,
however, that this is not a realistic description, as local Coulomb
repulsion (intraorbital $U$ and interorbital $U'$) inhibit $e_g$
electron motion for $x=0$ ($n=1$) limit, so the microscopic mechanism
of $E$-AF phase in HoMnO$_3$ remains puzzling.

\begin{figure}[t!]
\includegraphics[width=7.7cm]{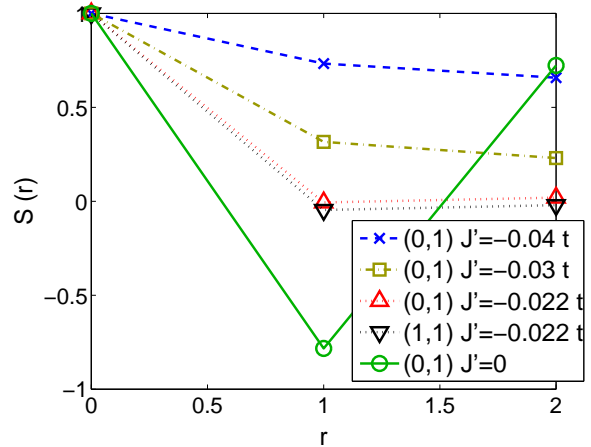}
\caption{(Color online)
Spin correlations ${\cal S}(\vec r)$ (\ref{ss}) in the
undoped monolayer as in Fig. \ref{fig:ss_MC_x0_2d}, but for decreasing
$J'$ from $J'=0$ (AF correlations) to $J'=-0.04t$ (FM correlations).
Parameters: $J=0.125t$, $E_z=-0.5t$, $\beta t=100$.
}
\label{fig:ss_MC_x0.5_2d}
\end{figure}

While ground state calculations comparing different ordered phases lead
to valuable insights at relatively low computational cost, one has to
realize that such calculations at $T=0$ are still necessarily limited
by the authors' imagination concerning the magnetic phases to consider.
We therefore complemented them by unbiased MC simulations for a few
parameter sets at low temperature $\beta t=100$. Figure
\ref{fig:ss_MC_x0_2d} shows the spin correlations (for core $t_{2g}$
spins) obtained from the MC data for one ($E_z=J'=0$) example of the FM,
and two ($E_z=0$, $J'=0.05t$ and $E_z=-0.5t$, $J'=0$) of the AF phase.
These data are complemented by the orbital correlations (not shown)
which correspond closely to the ground state results. Both spin and
orbital correlations weaken with rising temperature, as discussed
elsewhere,\cite{Dag05} but for the realistic values of $J'\simeq 0.02t$
spin correlations melt somewhat faster even in the present case when the
orbital interactions induced by the JT effect are neglected.

The situation is somewhat different for the $C$-AF phase obtained by
exact diagonalization at $T=0$ for $E_z\lesssim -0.2t$ (see Fig.
\ref{fig:phd1}). We analyzed the spin correlations ${\cal S}_{ij}$ at
low temperature for $E_z=-0.5t$ and a few selected values of $J'=0$,
$-0.022t$, $-0.03t$, $-0.04t$ (Fig. \ref{fig:ss_MC_x0.5_2d}). For $J'=0$
one finds AF correlations and for $J'=-0.04t$ FM ones, both in
accordance with the ground state phase diagram of Fig.~\ref{fig:phd1}.
However, the phase transition between these phases (the value
$J'=-0.022t$ lies right in the middle of the $C$-AF region) does not
occur via the $C$-AF phase at finite temperature, which would give an AF
signal for $\vec r=(1,1)$. Neither do the intermediate values of $J'$
exhibit the $E$-AF phase, which should show an AF signal for
$\vec r=(0,2)$ and has only slightly higher energy than the $C$-AF phase
at $T=0$, nor do the MC snapshots for this parameter range and
$\beta t=100$ suggest any other ordered phase. However, the temperature
$\beta t=100$ might be still too high to allow for longer range magnetic
correlations precisely in the crossover regime. In conclusion, we have
established that the AF and FM phases of Fig. \ref{fig:phd1} are well
supported by the MC results, while we believe the transition between
them might occur either via the $C$-AF phase, or with phase separation.
\cite{Dag03}

\begin{figure}[t!]
\includegraphics[width=7.7cm]{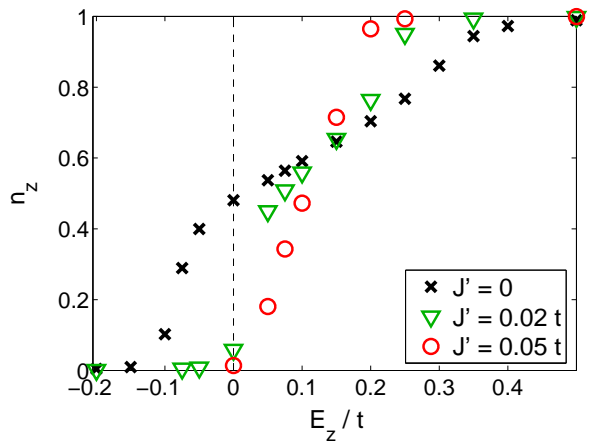}
\caption{(Color online)
Electron density in the $|z\rangle$ orbital perpendicular to an $ab$
plane of a monolayer manganite for increasing crystal field splitting
$E_z$, see Eq. (\ref{Hz}), as obtained from MC simulations of undoped
($n=1$) $\sqrt{8}\times\sqrt{8}$ clusters with various values of $J'$.
Parameters: $J=0.125t$, $\beta t=100$.
}
\label{fig:nz_vs_Ecr}
\end{figure}

In order to understand better the phase diagram of Fig.~\ref{fig:phd1}
it is instructive to consider the orbital occupation. When positive
crystal field ($E_z>0$) is applied, $|z\rangle$ orbitals which stick
out of the $ab$ plane are favored, see Fig.~\ref{fig:nz_vs_Ecr}. For
the AF phase at $J'=0.05t$, relatively small value $E_z=0.15t$ suffices
already to switch the density distribution from mainly $|x\rangle$ (at
$E_z=0$) to mainly $|z\rangle$ orbital occupation. The width of the
crossover regime from $|x\rangle$ to $|z\rangle$ orbital occupation
increases with decreasing $J'$. As discussed above, the FM phase is
characterized by an approximately equal occupation of both orbitals at
$E_z=0$. When both $e_g$ orbitals mix, the AO order found in the FM
phase at $J'=0$ is robust and a rather large value of crystal field
$E_z\simeq 0.4t$ is needed to enforce complete $|z\rangle$ polarization.
This large value of $E_z$ reflects the cooperative character of the FM
spin and AO order. At the same time, a negative crystal field
$E_z\lesssim -0.1 t$ is needed in order to enforce complete $|x\rangle$
polarization, while this transition occurs at $E_z\simeq 0$ for
$J'=0.02t$.

Note that in the ferro orbital (FO) phase with $|x\rangle$ orbitals occupied (at
$E_z<-0.2t$, the AF spin order is induced even for $J'=0$, see the spin
correlations in Fig.~\ref{fig:ss_MC_x0_2d} and the phase diagram of
Fig.~\ref{fig:phd1}. This provides another example of complementary
spin and orbital correlations that agree with Goodenough-Kanamori
rules.\cite{Goode} At $J'=0.02t$ one finds an intermediate situation
in this respect --- while both at $E_z\leq 0$ and $E_z\geq 0.25t$ the
the magnetic order is AF and dictated by $J'$, with either $|x\rangle$
or $|z\rangle$ orbitals being filled, in the transition region at
$0<E_z<0.25t$ one finds AO $(|x\rangle\pm|z\rangle)/\sqrt{2}$ order, but
rather unclear magnetic structure at $\beta t=100$. This shows again
that magnetic correlations are typically lost at lower temperature,
particularly when different conflicting trends in magnetic interactions
compete with each other.

Experimentally, the AF order with $e_g$ electrons occupying mainly
$|z\rangle$ orbitals was found\cite{Sen05} in a monolayer undoped
compound LaSrMnO$_4$, which we interpret as a consequence of
sufficiently large positive crystal field, $E_z\sim 0.5t$, induced by
the 2D structure. With growing temperature $|x\rangle$ occupation rises,
\cite{Sen05} as we also observed in the MC simulations.\cite{Dag05}

\subsection{Stability of the CE phase at half doping}
\label{sec:CE-mono}

\begin{figure}[t!]
\includegraphics[width=7.7cm]{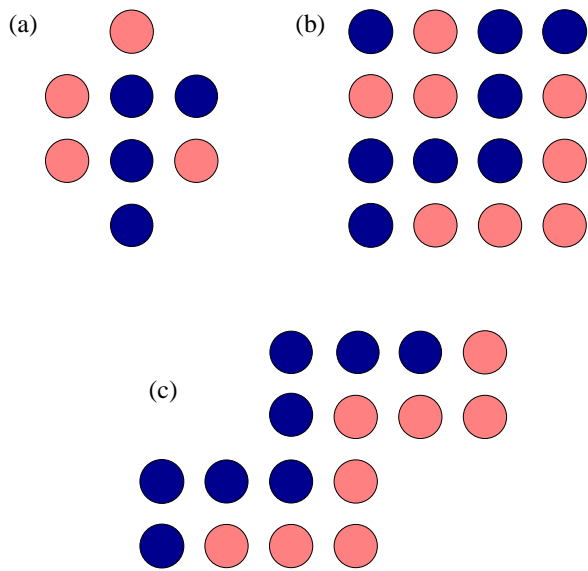}
\caption{(Color online)
Clusters used to investigate the CE phase at half doping ($x=0.5$):
(a) $\sqrt{8}\times\sqrt{8}$,
(b) $4\times4$, and
(c) $2\times8$.
Dark (light) shading indicates possible realizations of staggered
FM zig-zag chains in the CE phase.
}
\label{fig:CE_ideal}
\end{figure}

\begin{figure}[b!]
\includegraphics[width=7.7cm]{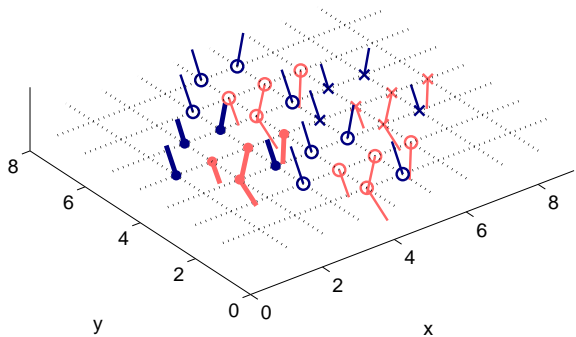}
\caption{(Color online)
MC snapshot obtained for doping $x=0.5$, with spin directions indicated
by lines. Four $\sqrt{8}\times\sqrt{8}$ clusters are shown (heavy lines
indicate the original cluster) and one clearly recognizes the magnetic
order in the CE phase.
Parameters: $J=0.125t$, $J'=0.05t$, $E_z=V=0$, $\beta t=100$.
}
\label{fig:ce_snapshot}
\end{figure}

Next we investigate the spin correlations at half doping ($x=0.5$) and
determine the range of stability of the CE phase, which was observed in
a monolayer La$_{0.5}$Sr$_{1.5}$MnO$_4$ compound.\cite{Ste96,Lar05} We
have performed ground state ($T=0$) calculations using three clusters
shown in Fig.~\ref{fig:CE_ideal}, with periodic boundary conditions.
Because of the large Hilbert space at half-filling, MC simulations could
be completed only for $\sqrt{8}\times\sqrt{8}$ clusters [for the cluster
geometry and one possible realization of the CE phase see Fig.
\ref{fig:CE_ideal}(a)]. In these simulations, we obtained the CE phase
for some parameters, e.g. for $J=0.125t$, $E_z=0$, either for
$J'=0.05t$ and $V=0$, or for $J'=0.025t$ and $V=t$. A typical MC
snapshot is shown in Fig. \ref{fig:ce_snapshot}. As we have used
periodic boundary conditions, one recognized the CE phase in an
$\sqrt{8}\times\sqrt{8}$ cluster repeated several times.

\begin{figure}[t!]
\includegraphics[width=7.7cm]{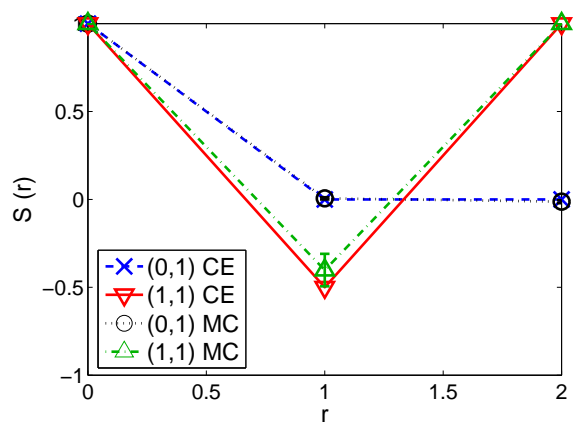}
\caption{(Color online)
Spin-spin correlation ${\cal S}(\vec r)$ (\ref{ss}) (\ref{ss}) for the
ideal CE phase (CE) and the MC data (MC) for an $\sqrt{8}\times\sqrt{8}$
cluster. There is one AF signal at $(1,1)$.
Parameters: $J=0.125t$, $J'=0.05t$, $E_z=V=0$, $\beta t=100$.
}
\label{fig:ss_CE}
\end{figure}

The spin correlations resulting from the MC runs are compared to those
of the ideal CE phase (exact diagonalization at $T=0$) in
Fig.~\ref{fig:ss_CE}, and one finds
an almost perfect agreement --- in both cases, AF and FM signals cancel
along $(01)$ direction, because of the different possible realizations
of the CE phase, but along the diagonal $(11)$ direction the spin
correlation reaches $-0.5$. (For the Monte Carlo simulations in these
parameter regimes, we used parallel
tempering\cite{Huk95} in order to sample the different symmetry-related
realizations correctly.)

On the one hand, the occurrence of the CE phase for classical spins even
\emph{without\/} the cooperative JT effect in the present study is in
contrast to the results obtained recently for quantum mechanical $S=1/2$
core spins,\cite{Bal04} where quantum fluctuations suppress the CE
phase. While it is not completely clear whether $S=3/2$ core spins are
better approximated by more quantum $S=1/2$ or by classical $S\to\infty$
spins, the latter has been shown to be an excellent approximation in the
one-orbital model in one dimension.\cite{Neu05} On the other hand,
a similar model, but including phonons and \emph{without\/} on-site
Coulomb repulsion between the $e_g$ electrons was treated on a
$4\times 4$ site cluster [see Fig. \ref{fig:CE_ideal}(b)] in Ref.
\onlinecite{Ali03}. Because the on-site Coulomb repulsion was not
included, unrealistically large $J'>0.2t$ or rather strong
electron-phonon coupling $\lambda\gtrsim 1.75$ were necessary to
stabilize the CE phase. Although this result could still be improved by
considering larger clusters, it highlights already the importance of
strong Coulomb interactions for obtaining physically relevant results.

When FM and AF interactions occur simultaneously at different bonds in
$ab$ planes, one has to investigate the stability of the CE phase by
comparing it with the $C$-AF phase which has the same amount of FM and
AF bonds. We have found that at $J=0.125t$ the $C$-AF phase has higher
energy than the CE one for small clusters of eight sites, and the
same holds true for larger $4\times 4$ clusters. However, it has been
pointed out that finite size effects are important,\cite{She01} so one
would like to investigate still larger clusters, at least at $T=0$.

As a first step, we exploit the quasi-1D
nature of the two phases and investigate $8\times 2$
clusters instead of $4\times 4$ ones (at present, we cannot treat
more than 16 sites):
 (i) a ladder for the $C$-AF phase, and
(ii) the cluster shown in Fig.~\ref{fig:CE_ideal}(c) for the CE phase.
While the energy of the CE phase hardly changes with cluster topology,
the one of the $C$-AF phase is considerably lowered, albeit still higher
than that of CE phase, see Tab.~\ref{tab:e_c_ce}. To
make the ground state calculations more conclusive and to eliminate
systematic errors, it is therefore indeed necessary to investigate
finite size effects.

Due to the magnetic order in both $C$-AF and CE phases, hopping occurs
only along a 1D path, so chains can be investigated instead of 2D
lattices. In this way larger systems could be reached. Taking the
kinetic energy for a FM chain,
\begin{eqnarray}
H_t^{\rm C/CE} &=& -\frac{1}{4}t\sum_{i}
  \big[3{\tilde c}_{i,x}^{\dagger}{\tilde c}_{i+1,x}^{}
       +{\tilde c}_{i,z}^{\dagger}{\tilde c}_{i+1,z}^{}     \nonumber \\
 &&  \pm\sqrt{3}
 ({\tilde c}_{i,x}^{\dagger}{\tilde c}_{i+1,z}^{}
 +{\tilde c}_{i,z}^{\dagger}{\tilde c}_{i+1,x}^{}) + \mathrm{h.c.}\big],
\label{Ht_CE}
\end{eqnarray}
we considered two different geometries:
 (i) the $C$-AF phase with a chain along $b$ axis, i.e., taking $+$
     sign for the interorbital hopping in Eq. \ref{Ht_CE}, and
(ii) a zig-zag path chosen instead for the CE phase, which leads to a
     sequence $(b,b,a,a,b,b,\dots)$ of bond directions and therefore
     to the phase sequence $(+,+,-,-,+,+,\dots)$ in the hopping.

Because the FM order within the chains (in either CE or $C$-AF phase)
is perfect at zero temperature ($u_{ij}=1$), the superexchange along
them contains only one term,
\begin{equation}
\label{HFM}
H_J^{\rm FM}= J\sum_{i}
\Big( 2T_{i}^{\zeta}T_{i+1}^{\zeta}
-\textstyle{\frac{1}{2}{\tilde n}_i{\tilde n}_{i+1}} \Big)\;,
\end{equation}
where $\zeta$ is the directional orbital along the bond direction, i.e.,
along $b$ in the $C$-AF phase and alternating between $a$ and $b$ in the
CE phase. However, superexchange via AF bonds also contributes, and these
\emph{interchain coupling} terms could be crucial for stabilizing one
or the other phase.\cite{Bri01r} We therefore embedded the chains and
included into the effective 1D Hamiltonian additional AF superexchange
terms,
\begin{eqnarray}
\label{HJAF}
H_J^{\rm AF}&=& \frac{3}{5}J\sum_{i}
\Big( 2T_{i}^{\bar\zeta}T_{i+1}^{\bar\zeta}
-\textstyle{\frac{1}{2}{\tilde n}_i{\tilde n}_{i+1}} \Big) \nonumber \\
&-&J\sum_{i}\big[{\tilde n}_{i\bar\zeta}(1\!-\!{\tilde n}_{i+1})
+(1\!-\!{\tilde n}_i){\tilde n}_{i+1\bar\zeta}\big]        \nonumber \\
&-&\frac{9}{10}J\sum_i{\tilde n}_{i\bar\zeta}{\tilde n}_{i+1\bar\zeta}\;,
\end{eqnarray}
where $\bar\zeta$ is the directional orbital \emph{perpendicular} to
the chosen path, i.e., couples nearest neighbor sites on two adjacent
chains. The form of Eq. (\ref{HJAF}) is motivated by the fact that
a bridge site on one chain (in CE phase) lies always next to two corner
sites of the neighboring chains in $ab$ plane. By symmetry, these should
be equivalent to the corner sites on the considered chain, which are
again the nearest neighbor sites to a given bridge site. The energy of
the $C$-AF phase is evaluated with a similar term.

\begin{table}[b!]
\caption{
Ground state energies as obtained for $C$-AF and CE phases with
different clusters of Fig. \ref{fig:CE_ideal} and for 1D chains
simulating these phases, for three representative values of $E_z$.
In case of $2\times 8$ clusters, a ladder was used for the $C$-AF phase
and the cluster shown in Fig.~\ref{fig:CE_ideal}(c) for the CE phase.
Parameters: $J=0.125t$, $V=0$.
}
\label{tab:e_c_ce}
\vskip .1cm
\begin{ruledtabular}
\begin{tabular}{ccccccc}
$E_z/t$ &phase &$\sqrt{8}\times\sqrt{8}$&$4\times 4$&$2\times 8$
                                        &$4\times1$ &$8\times1$\cr
\colrule
 0.0 &  CE    &-0.6452 &-0.6624 &-0.6624 &-0.6466 &-0.6641 \cr
     & $C$-AF &-0.5366 &-0.5366 &-0.6330 &-0.5366 &-0.6365 \cr
\colrule
 0.5 & CE     &-0.7371 &-0.7358 &-0.7358 &-0.7372 &-0.7357 \cr
     & $C$-AF &-0.6220 &-0.6221 &-0.7068 &-0.6140 &-0.7076 \cr
\colrule
-0.5 & CE     &-0.5895 &-0.6238 &-0.6238 &-0.5917 &-0.6269 \cr
     & $C$-AF &-0.5183 &-0.5184 &-0.6045 &-0.5159 &-0.6077 \cr
\end{tabular}
\end{ruledtabular}
\label{tab:ce}
\end{table}

In order to check the validity of this one-dimensional approach, we
compare the energies obtained on small 2D clusters (third to fifth
column in Tab.~\ref{tab:ce}) to those obtained on short chains of four
and eight sites (last two columns in Tab.~\ref{tab:ce}).
For the CE phase, $L=4$ corresponds to the
chain length encountered in a $\sqrt{8}\times\sqrt{8}$ cluster, and the
energies differ indeed only by $\sim 0.0015t$. For the $C$-AF phase,
$L=4$ corresponds to either $\sqrt{8}\times\sqrt{8}$ or $4\times4$
cluster, and one also finds remarkably good agreement. The difference is
larger for $L=8$ and the ladderlike clusters, but it seems still
reasonable to investigate systematically the Hamiltonian comprising the
three terms given in Eqs. (\ref{Ht_CE}), (\ref{HFM}), and (\ref{HJAF})
on chains of different length.

\begin{figure}
\includegraphics[width=8.0cm]{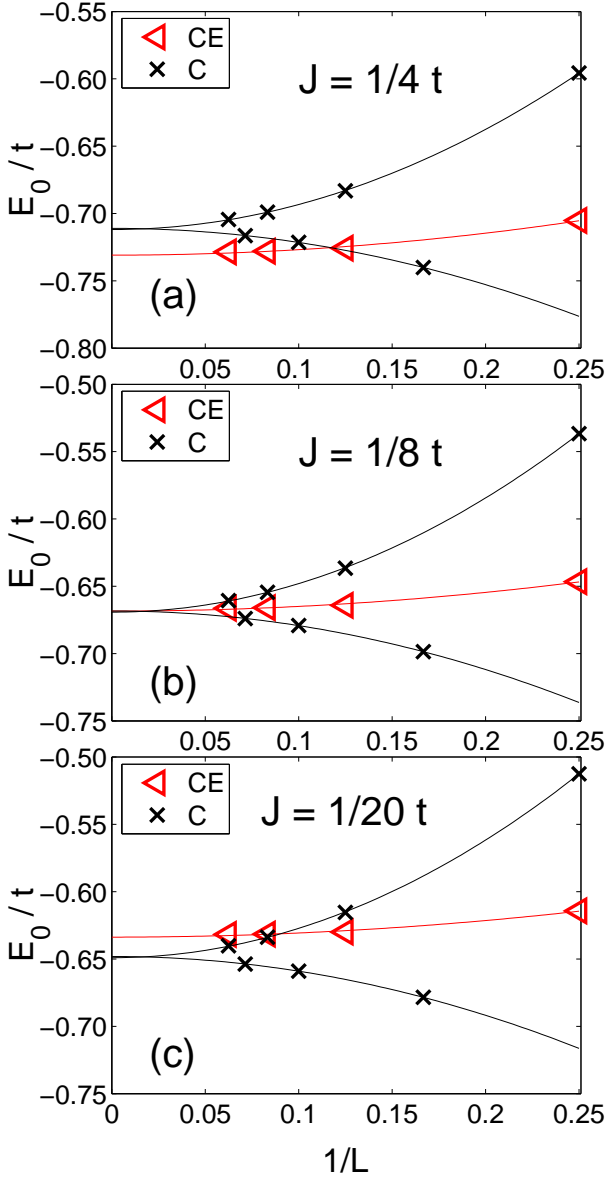}
\caption{(Color online)
Finite size extrapolation of the ground state energy $E_0$ for the
$C$-AF and CE phases on chains of different length $L\leq 16$, as
obtained at $T=0$ for different values of $J$:
(a) $J=0.25t$,
(b) $J=0.125t$, and
(c) $J=0.05t$.
The data for the $C$-AF phase alternate between the chains of length
$L=4n$ and $L=4n+2$, with $n$ integer.
Parameters: $J'=E_z=V=0$.
}
\label{fig:cce}
\end{figure}

The results for chains of various lengths were next extrapolated to the
$1/L\to 0$ limit (see Fig. \ref{fig:cce}). An excellent fit was obtained
using a quadratic dependence of the ground state energy on the inverse
chain length, $E_0=E_{0,\infty}+k\cdot\frac{1}{L^2}$. In this way we
deduced the extrapolated energy values, $E_{0,\infty}$. As only
directional orbitals oriented along the chain contribute to the kinetic
energy in the $C$-AF phase, this energy is not influenced by $U$.
Consequently, one finds that large on-site repulsion $U$, i.e., small
$J$, favors the $C$-AF phase (see also Refs. \onlinecite{She01,Bri01r}).
Furthermore, it appears that the energies of both phases are so close to
each other for $J\sim t/8$ that one cannot distinguish between these
phases and decide on the nature of magnetic correlations in the ground
state. This result is not strongly affected by a uniform crystal field
$E_z$ either --- the energies of the two magnetic phases are again very
similar for $J=t/8$.

As the commonly used picture of the CE phase assumes charge order, one
expects that it to be stabilized by nearest neighbor Coulomb repulsion
$V$. The extrapolation to $1/L\to 0$ gives indeed lower energies of the
CE phase for $V>0$, but the effect of $V$ remains \emph{surprisingly
small}. The reason is that the second of the AF terms in Eq.
(\ref{HJAF}) already induces some charge order in the $C$-AF phase as
well, and therefore its energy does not suffer much from Coulomb
repulsion $V$. If, on the other hand, $V$ becomes too large, it hinders
electron motion along the FM chains in both the $C$-AF and CE phases and
thus affects both of them, even favoring the $C$-type AF phase for very
large $V\gtrsim 1.5t$.

\subsection{Orbital and charge order at half doping}
\label{sec:oo}

When the magnetic order of CE type occurs, the sites along each FM
zig-zag chain are nonequivalent, and one expects that holes are
predominantly found at the corner sites.\cite{Dag03,Hot03,Cuo02}
However, in the present finite cluster calculations different CE
patterns mix with each other and the holes cannot be detected using
just density operators. This information can be extracted only from
intersite correlations, such as for instance the
spin-hole-spin correlation ${\cal R}$ which measures the spin-spin
correlations across a central site occupied by the hole, see Eq.
(\ref{sns}). For the clusters shown in Fig.~\ref{fig:CE_ideal} we
consider the FM zig-zag chains along (11) direction. On the one hand,
the sites to the `right' and `left' of a corner site along $a$ axis,
i.e., (10) direction, have opposite spin, as well as those `above' and
`below' it along $b$ axis, i.e. (01) direction. For the bridge sites,
on the other hand, the neighboring spins along either direction have
the same sign. In the CE phase, negative values of ${\cal R}$ indicate
therefore that holes occupy corner rather than bridge sites.
\cite{notedi}

We have calculated the correlation function ${\cal R}$ for the ground
state, i.e., first the spins were set into the zig-zag CE pattern and
next the resulting orbital Hamiltonian was solved with Lanczos
diagonalization. In the absence of the nearest neighbor Coulomb
repulsion (at $V=0$), we then found ${\cal R}=-0.087$, from which we
deduced the electron density at corner sites $n_c\simeq 0.413$ vs.
$n_b\simeq 0.587$ on the bridge sites. (The MC data at low temperature
$\beta t=100$ with $J'=0.05 t$ and $E_z=V=0$ give a somewhat weaker
spin/charge order with ${\cal R}=-0.0757\pm 0.001$.) The electrons at
the bridge sites are almost exclusively in the directional $3x^2-r^2$
($3y^2-r^2$) orbitals along $a$ ($b$) axis, i.e., along the direction of
the zig-zag chain. In contrast, the electrons at corner sites are more
evenly distributed over both orbitals in the CE ground state
(at $E_z=0$), with $n^c_x=0.2307$ in $|x\rangle$ vs. $n^c_z=0.1823$ in
$|z\rangle$ orbital, respectively.

\begin{figure}[t!]
\includegraphics[width=8.2cm]{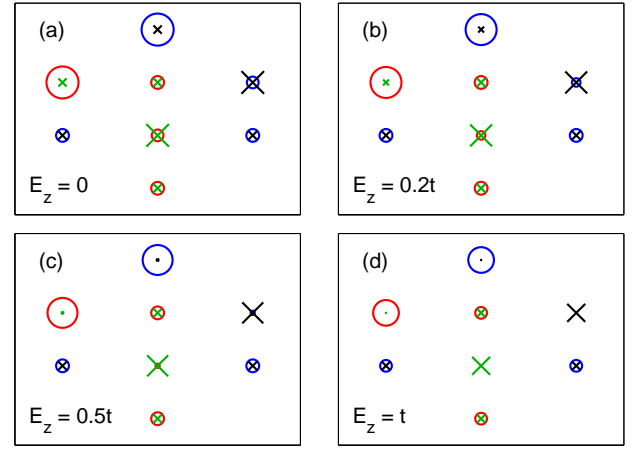}
\caption{(Color online)
Orbital structure for the CE phase at $T=0$, with circles (crosses)
for $3x^2-r^2$ ($3y^2-r^2$) orbitals, as obtained for:
(a) $E_z=0$,
(b) $E_z=0.2t$,
(c) $E_z=0.5t$, and
(d) $E_z=t$.
The size of circles and crosses is proportional to the electron density
in a given orbital. Parameters: $J=0.125t$, $V=0$.
}
\label{fig:z_y_Ezvar}
\end{figure}

These findings, i.e., the relatively small difference in the density at
bridge and corner sites and the occupation of the directional orbitals,
are in contrast to some experimental results.\cite{Hua04,Wil05}
However, the OO similar to our findings was also reported,
\cite{Dhe04,Oga01,Yam00} and one expects that rather extreme charge
modulation, with holes at corner sites and $e_g$ electrons in bridge
positions, should be excluded as then the FM double exchange which
stabilizes the CE phase would be lost. Therefore, the charge order
with alternating Mn$^{3+}$ and Mn$^{4+}$ ions has been challenged, and
intermediate valence picture with only small density variations has been
suggested both in experimental\cite{Mar05} and in theoretical studies.
\cite{Ali03,Mah01,Cuo02,Eba05}

In this Section we would like to address as well the recent controversy
concerning the type of the OO in the CE phase. We analyzed the orbital
occupation in the $3x^2-r^2$ and $3y^2-r^2$ orbitals for various values
of the crystal field splitting $E_z\geq0$ (Fig. \ref{fig:z_y_Ezvar}).
At $E_z=0$ the electrons at the bridge sites are found in the orbitals
parallel to the FM chain [Fig. \ref{fig:z_y_Ezvar}(a)], i.e., in
$3x^2-r^2$ or $3y^2-r^2$, as discussed above. One finds that the
corresponding orthogonal orbital is practically empty, e.g. $y^2-z^2$
orbitals at the bridge sites with both neighboring FM bonds along $x$
direction. However, large density is found within these orbitals at the
remaining bridge positions (in $y^2-z^2$ orbitals on those sites where
the FM chains have the bonds along $y$ direction), which is due to the
considerable overlap between the $3y^2-r^2$ and $y^2-z^2$ orbitals.
Thus, the $3x^2-r^2$/$3y^2-r^2$ order appears to be qualitatively
similar to the $z^2-x^2/y^2-z^2$ order. In the present case, however,
we would rather identify it as $3x^2-r^2$/$3y^2-r^2$ order because the
occupation on the bridge sites is almost exclusively in the directional
orbitals.

\begin{figure}[t!]
\includegraphics[width=8.5cm]{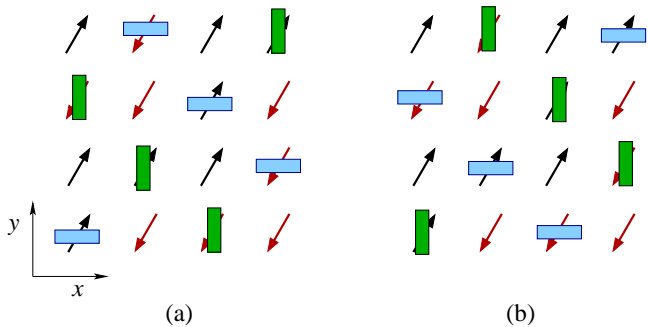}
\caption{(Color online)
Possible orbital and spin structure for the CE phase, as suggested:
(a) in Fig. 1 of Ref.~\onlinecite{Wil05}, and
(b) from our calculations.
The orientation of the rectangles indicates the type of occupied
orbital --- horizontal $z^2-x^2$ or vertical $y^2-z^2$.
}\label{fig:CE_orbs}
\end{figure}

When the crystal field $E_z$ increases, the orbital occupation changes
(see Fig.~\ref{fig:z_y_Ezvar}). Not surprisingly, density in the
$3z^2-r^2$ orbitals (not shown) has increased, which modifies the type
of the OO --- one finds now that the \emph{directional} orbital is the
one which is empty on some sites, which means that the electron on a
bridge site has moved from the $3x^2-r^2$ ($3y^2-r^2$) into the
$z^2-x^2$ ($y^2-z^2$) orbital. This situation could therefore be
considered as representing $z^2-x^2/y^2-z^2$-type order. The observed
transition from the former (at $E_z=0$) to the latter (at $E_z=1$)
phase is driven by the crystal field and is gradual, see Fig.
\ref{fig:z_y_Ezvar}. Therefore, it may be argued that the combination
of spin and orbital structure in Fig.~1 of Ref. \onlinecite{Wil05},
given schematically in Fig.~\ref{fig:CE_orbs}(a), is incorrect: In this
picture, the FM spin chains run perpendicular to the occupied $z^2-x^2$
($y^2-z^2$) orbitals at the bridge sites, i.e., in $y$-($x$-) direction.
Instead, our data indicate that the FM chains run as shown in Fig.
\ref{fig:CE_orbs}(b), i.e., in $x$-direction, where the $z^2-x^2$
orbital is occupied for large $E_z>0$ and along $y$-direction for the
$y^2-z^2$ sites.

Another difference between the analysis performed in Refs.
\onlinecite{Wil05} and our results is that charge order is not perfect
in our case, as it was assumed in their analysis. This does not
influence the OO, however --- for $V=t$ the charge order is enhanced
(i.e. the electron density at bridge sites increases) without greatly
affecting the type of the OO (not shown). Here we would like to
emphasize that our findings concerning the nature of spin and OO,
as well as rather weak charge order, agree with the recent
Hartree-Fock calculations on the multiband $d$-$p$ model,\cite{Eba05}
indicating that the local Coulomb interactions and superexchange suffice
already to stabilize the CE phase. There is no doubt, however, that
the oxygens distortions also contribute to the stability of this phase,
\cite{Bal04,Eba05} and would expand the regions of the CE phase in the
phase diagrams shown in the next Section. At the same time realistic
oxygens distortions would also modify somewhat the type of occupied
orbitals, but the essential features of the orbital order in the CE
phase, see Fig.~\ref{fig:CE_orbs}(b), would remain the same.

The facts that $z^2-x^2/y^2-z^2$- and $3x^2-r^2$/$3y^2-r^2$-type orbital
order are qualitatively similar,\cite{notece} and that one can come from
one to the other one by adding a constant crystal field, are consistent
with the results reported in Ref.~\onlinecite{Hua04}, where a shear-type
distortion (alternating \emph{contractions} along $x/y$-axes) has been
found more plausible than a JT type OO (alternating \emph{elongations}
along $x$/$y$-axes). In the shear-type order, the out-of-plane Mn--O
bond is of similar length as the \emph{longer} in-plane bond, while it
is comparable to the \emph{shorter} bond in the JT-like case. Variation
of the out-of-plane bond, equivalent to $E_z>0$ in our model, can
therefore lead from one scenario to the other. In closing, we remark
that the kinetic energy favors $3x^2-r^2$/$3y^2-r^2$--occupation as this
maximizes the hopping. Onsite Coulomb repulsion inhibits the kinetic
energy and when we weaken Coulomb repulsion from $U=8 t$ ($J= 0.125 t$)
to $U=4t$ ($J=0.25t$) and thus enhance the kinetic energy, we indeed
find that stronger $E_z$ is needed to induce $z^2-x^2/y^2-z^2$-type OO.
This is in accordance with the usual experience based on comparing LDA
with LDA+U calculations,\cite{Hua04} where inclusion of onsite
interaction has likewise been found to be crucial in stabilizing
shear-type over JT type OO.

\subsection{Phase diagrams at half doping}
\label{sec:phd}

\begin{figure}[t!]
\includegraphics[width = 7.7cm]{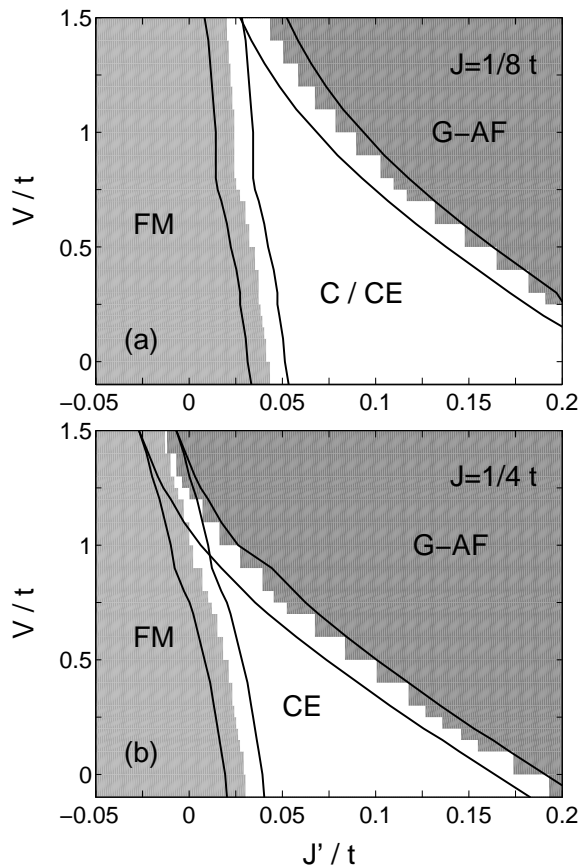}
\caption{
Phase diagram at half doping ($x=0.5$) for:
(a) $J=0.125t$, and
(b) $J=-.25t$,
as obtained from finite size considerations at orbital degeneracy
($E_z=0$). The black lines around the phase boundaries give their
estimated numerical and finite-size errors. At $J/t=1/8$ the energies for the $C$-AF
and CE phases differ only \emph{very little} and we could not determine
a phase boundary between them. For $J/t=1/4$ the CE phase has lower
energy than the $C$-AF phase.
}
\label{fig:phases_fs}
\end{figure}

Our results on the stability of magnetic phase at $x=0.5$ were collected
in the phase diagrams of Fig. \ref{fig:phases_fs} for two representative
values of the on-site Coulomb repulsion, $U=8t$ and $4t$, leading to
$J=t/8$ and $t/4$.
For a more realistic $J=t/8$, the extrapolated energies for the $C$-AF
and CE phases are indeed very close to each other, with $V>0$ slightly
favoring the CE phase (see above), and we anticipate that the observed
differences might actually be smaller than the error induced by the use
of 1D chains instead of 2D clusters. We therefore could not determine a
phase boundary between these two phases. We also had to compare the
energies of the two quasi-1D phases to the 2D-like FM and AF phases.
However, we observed that the energy of the CE phase for $L=8$ is
already very similar to the extrapolated result for $L\to\infty$ for
all values of $V$. Therefore the results obtained for the $8\times2$
cluster, see Fig. \ref{fig:CE_ideal}(c), could be used for energy
comparison with the FM and AF phases.

In order to estimate the uncertainty of the ground-state energies in a
reliable way, and thus of the determined phase boundaries, we further
investigated the change of the energy of the FM phase with the cluster
size for three clusters used in the present calculations:
$\sqrt{8}\times\sqrt{8}$, $\sqrt{10}\times\sqrt{10}$, and $4\times4$.
It was found that the energies did not depend much on cluster size,
but these might, however, still be too small. A somewhat simpler
situation occurs for the $G$-type AF phase which turns out to be
perfectly charge ordered for $V\gtrsim -J$, so its energy is independent
of $V$. Moreover, as all electrons are perfectly localized, one finds in
the present approximation a classically ordered $G$-AF phase without
finite-size effects.

For the $C$-AF and CE phases, we investigated how much the energy of the
CE phase for $L=8$ differs from the extrapolated energies for either the
$C$-AF or the CE phase at $L\to\infty$. We finally arrived at the
estimated error $\Delta J'\leq\frac{\Delta E}{2}\approx 0.01t$ and the
resulting phase diagram for $J/t=1/8$ given in Fig.
\ref{fig:phases_fs}(a). The CE or $C$-AF phases are stable in between
the AF and FM phases, respectively. Increasing nearest-neighbor Coulomb
repulsion $V>0$ weakly suppresses the FM phase and favors the AF phase.
On the one hand, the FM phase is suppressed because it is stabilized by
the kinetic energy which decreases when the charge order is induced by
finite $V$. The AF phase, on the other hand, is already charge ordered
in the absence of explicit Coulomb repulsion and is therefore not
influenced by $V$. Altogether, the electrons redistribute at finite $V$
also in the CE phase, so its range shrinks.

For a smaller on-site Coulomb repulsion $U=4t$, e.g. $J/t=1/4$, the CE
phase is favored over the $C$ phase for all values of $V$, at least in
the investigated parameter range $-0.5t\leq V\leq 2t$. We used the same
procedure as described above to determine the boundary toward the FM and
AF phases, and arrived at the phase diagram depicted in Fig.
\ref{fig:phases_fs}(b). One finds that the $G$-AF phase extends to a
broader range of parameters, and the region of CE phase is reduced with
increasing $V$. This contradicts the common belief that the CE phase is
stabilized by the nearest neighbor Coulomb interaction.

\subsection{Magnetic and orbital correlations for increasing doping }
\label{sec:incr_doping}

We complete this Section by a qualitative discussion of the changes in
orbital occupation and magnetic correlations in a monolayer under
increasing doping. For a $4\times4$ cluster doped with one hole
($x=1/16=0.0625$) we observed rather weak AF order in MC runs for
$J'=0.02t$ and more pronounced AF order for $J'=0.05t$, with $E_z=V=0$.
Unfortunately, we were not able to perform MC simulations for more than
one hole on $4\times4$ clusters ($x>1/16=0.065$) due to the
increasing size of the Hilbert space. For one hole doped to a smaller
$\sqrt{8}\times\sqrt{8}$ cluster (doping $x=1/8=0.125$), we found weakly
AF correlations for $J'=0.05t$, while AF order had vanished for
$J'=0.02t$, which is a clear sign of the increasing importance of FM
double exchange with increasing doping. The fast disappearance of AF
order ($x\gtrsim 0.125$ for $J'=0.02 t$) agrees with experimental
observations, where AF order disappears at $x\gtrsim0.115$, and is
replaced by short-range spin-glass type of order.\cite{Lar05}

Magnetic correlations at higher doping are theoretically challenging,
and various anisotropic magnetic phases ($A$-AF and $C$-AF) were
obtained in the model discarding strong Coulomb repulsion.\cite{Bri99}
In the range of large doping $x>0.5$, higher doping can be reached for
Nd$_{1-x}$Sr$_{1+x}$MnO$_4$, while La$_{1-x}$Sr$_{1+x}$MnO$_4$ can be
doped only up to $x\approx 0.7$. In Nd$_{1-x}$Sr$_{1+x}$MnO$_4$ samples
the $C$-AF phase was observed\cite{Kim02} for $0.75<x<0.9$;
additionally, a structural phase transition suggesting predominant
occupation of directional orbitals along one axis was reported. We
found it very encouraging that the same trends have been observed in
the MC simulations for two electrons on a $\sqrt{8}\times\sqrt{8}$
cluster and for four electrons on a $4\times 4$ cluster ($x=3/4$), as
well as for three electrons on $4\times 4$ cluster ($x=0.8125$), for
large enough $J'=0.025t$ and $0.05t$.

\begin{figure}[t!]
\includegraphics[width=8.0cm]{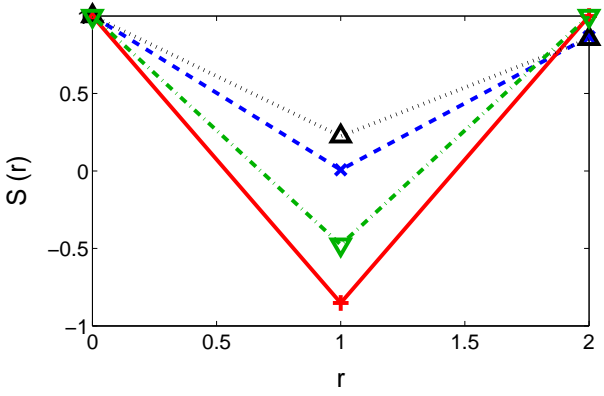}
\caption{(Color online)
MC results for the spin-spin correlations ${\cal S}(\vec r)$ (\ref{ss})
for a monolayer, as obtained in $\sqrt{8}\times\sqrt{8}$ cluster with
two electrons (at high $x=0.75$ doping) for two parameter sets:
$J'=0.05 t$, $V=0$, $E_z=0$ [$\times$ for $(10)$ and $+$ for $(11)$
                             direction];
$J'=0.025t$, $V=t$, $E_z=0.25t$ [$\triangle$ for $(10)$ and
                                 $\triangledown$ for $(11)$ direction].
Error bars are smaller than symbol sizes.
Parameter: $J=0.125t$, $\beta t =100$.
}
\label{fig:ss_2D_C_2e}
\end{figure}

The spin correlations obtained from MC simulations of a
$\sqrt{8}\times\sqrt{8}$ cluster with two electrons are shown in Fig.
\ref{fig:ss_2D_C_2e} for two parameter sets: $J'=0.05t$, $V=E_z=0$, and
$J'=0.025t$, $V=t$, $E_z=0.25t$. For a larger value of $J'=0.05t$ we see
the telltale signal of the $C$-type phase, the strongly negative signal
at $(11)$. The nearest-neighbor spin correlation in $(10)$ direction
nearly vanishes because the FM and AF signals from the two directions
almost cancel each other. (Parallel tempering was
employed in this case to ensure good autocorrelation times.) For smaller
$J'=0.025t$, and $V=t$, $E_z=0.25t$, the $C$-AF phase is not as marked,
and the FM correlations are stronger than the AF ones. In agreement with
expectations based on the 1D model,\cite{Dag04} the electrons occupy
directional orbitals along the FM direction in the $C$-AF phase, because
this maximizes the gain of the kinetic energy.

\begin{figure}[t!]
\includegraphics[width=8.0cm]{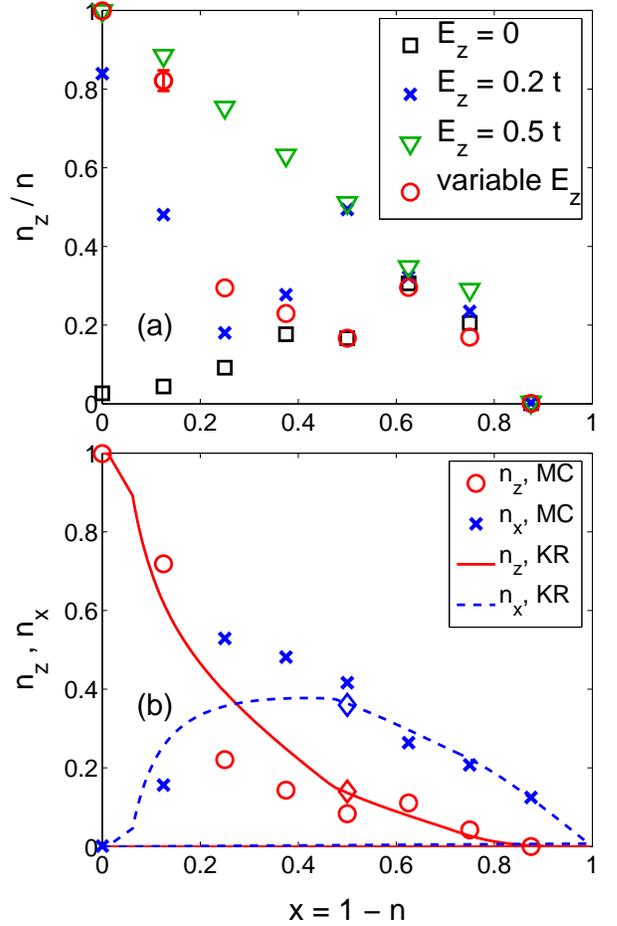}
\caption{(Color online)
Orbital electron densities for increasing doping $x=1-n$ as obtained in
MC method for a $\sqrt{8}\times\sqrt{8}$ cluster:
(a) percentage of density in $|z\rangle$ orbitals $n_z/n$ for a few
selected values of crystal field parameter $E_z$ and for variable $E_z$
(see below);
(b) absolute densities $n_z$ and $n_x$ in the two orbitals from
MC (symbols) compared with the densities found with the KR slave boson
mean-field approximation\cite{Kot86} (lines) with variable
$E_z=(\frac{1}{2}-x)t$.
The charge distribution found at doping $x=0.5$ with the KR method
($n_z\simeq 0.14$, $n_x\simeq 0.36$) is indicated by diamonds.
Remaining parameters in MC calculations:
$J=0.125t$, $J'=0.025t$, $V=t$, $\beta t=100$.
}
\label{fig:nmc}
\end{figure}

Figure \ref{fig:nmc} shows that orbital occupation depends strongly on
doping for realistic parameters, $J'=0.025t$ and $V=t$. This is
reflected by the percentage of electrons occupying out-of-plane
$|z\rangle$ orbitals which is furthermore very
sensitive to the actual value of $E_z$, see Fig. \ref{fig:nmc}. It is
quite remarkable that for the degenerate $e_g$ orbitals, i.e., without
crystal field ($E_z=0$), practically only in-plane $|x\rangle$ orbitals
are occupied at $x=0$ [squares in Fig. \ref{fig:nmc}(a)]. This state is
induced by finite AF superexchange $J'=0.025t$ which (due to large
Hund's exchange) selects the AF interactions between $e_g$, and these
interactions are maximal in $ab$ planes when $|x\rangle$ orbitals are
occupied. In this way the core spin superexchange influences also the
$e_g$ orbital occupation (at $J'=0$ one finds an almost isotropic
electron distribution with $n_z/n\simeq 0.5$ in the FM phase). Upon
doping, electron population in $|z\rangle$ orbitals gradually increases,
reaches a maximum at $x=5/8$ where FM correlations are found, and then
decreases again. This behavior follows from the kinetic energy which
contributes in the entire regime of $0<x<1$, and is gained when the
interorbital processes which excite electrons to $|z\rangle$ orbitals,
$\sim\sqrt{3}({\tilde c}_{ix}^{\dagger}{\tilde c}_{jz}^{}
+{\tilde c}_{iz}^{\dagger}{\tilde c}_{jx}^{})/4$, are allowed.

For a large positive crystal field favoring $|z\rangle$ orbitals
$E_z=0.5t$ [triangles in Fig. \ref{fig:nmc}(a)], the density
distribution is reversed at $x=0$ --- almost all electrons are found
within $|z\rangle$ orbitals. They redistribute, however, gradually with
increasing doping, because the kinetic energy competes with the crystal
field and favors in-plane $|x\rangle$ orbitals. (In this case one finds
a FM phase in a range of doping from $x=1/4$ to $x=5/8$, separated by
the CE phase found at $x=0.5$ for $E_z=0$.) In the intermediate case of
a smaller crystal field splitting of $E_z=0.2 t$ (crosses), the
normalized $|z\rangle$ density $n_z/n$ is first reduced but next rises
again towards the FM phase which in this case suppresses the CE phase
at $x=0.5$.

As the last scenario, we investigated the effect of a doping-dependent
crystal-field splitting,
\begin{equation}
E_z=\Big(\frac{1}{2}-x\Big)t.
\label{Ezx}
\end{equation}
The doping dependence of this type [circles in Fig.~\ref{fig:nmc}(a)]
is qualitatively expected by considering the experimental data.
\cite{Sen05} This is probably the most realistic case of those
considered here, as it shows both the correct orbital polarization for
$x=0$ and the CE phase for $x=0.5$. Note that in all cases except
$E_z=0$, we find the experimentally observed\cite{Sen05} increase of
in-plane $|x\rangle$ electron density with increasing doping which shows
that $E_z>0$ in the low doping regime.

Taking the decreasing with $x$ crystal field as in Eq. (\ref{Ezx}),
one finds a very fast decrease of $n_z$ from $n_z=1$ at $x=0$ to
$n_z\simeq 0.22$ at $x=0.25$ [Fig.~\ref{fig:nmc}(b)]. Thus the electrons
move fast from $|z\rangle$ to $|x\rangle$ orbitals in this doping
regime, as dictated by the kinetic energy gain. This qualitative trend
is very well reproduced by the analytic approach using the mean-field
approximation in the slave-boson method, introduced by Kotliar and
Ruckenstein.\cite{Kot86} We have adapted this method to the FM phase in
a monolayer, as explained in the Appendix. The performed comparison with
the results of exact diagonalization shows that this analytic approach
provides a surprisingly reliable way to estimate both the charge
distribution in layered systems and the magnetic interactions at
increasing doping. In the present case the double exchange mechanism
promotes FM states in a large range of doping, while in a bilayer
system changing electron density distribution provides a natural
explanation for an observed transition from the FM to $A$-AF structure.
\cite{Ole03}

\begin{figure}[t!]
\includegraphics[width=8.0cm]{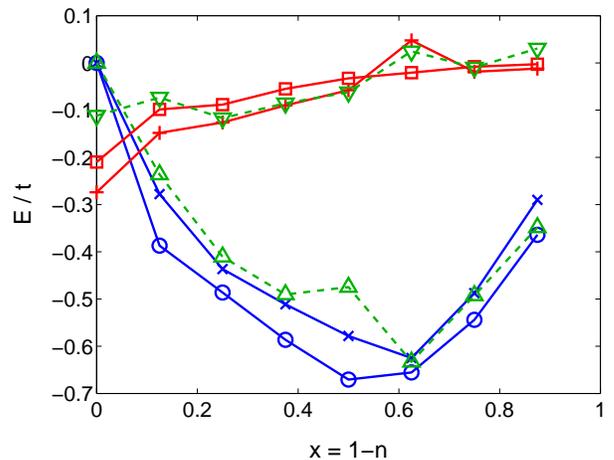}
\caption{(Color online)
Kinetic (approximately parabolic shape) and superexchange (weakly
increasing) energies for increasing doping $x=1-n$, as obtained for
$\sqrt{8}\times\sqrt{8}$ clusters with various parameter sets:
$\bigcirc$ and $\Box$ --- for $J'=0$, and $E_z=V=0$;
$\times$ and $+$ --- for $J'=0.05t$, and $E_z=V=0$;
$\triangle$ and $\triangledown$ --- for $J'=0.025t$, $V=t$,
and $E_z$ given by Eq. (\ref{Ezx}).
Parameters: $J=0.125t$, $\beta t=100$.
}
\label{fig:e_x}
\end{figure}

The changes in magnetic correlations at increasing doping follow from
the competing superexchange and double exchange interactions. The double
exchange is directly proportional to the kinetic energy,\cite{Ole02}
which vanishes at $x=0$ and is gradually gained by doping when the
available space for hopping processes increases up to $x\sim 0.5$, and
then is gradually lost beyond half doping (see Fig.~\ref{fig:e_x}).
Therefore, the kinetic energy has an approximately parabolic shape, as
obtained for a $\sqrt{8}\times\sqrt{8}$ cluster with a few
representative parameter sets. Since different orbital occupation can
favor either FM or AF spin configuration for any bond, the superexchange
energy is finite even in the cases where the average nearest-neighbor
spin correlation function vanishes, like in the CE and $C$-AF phases.

In the FM case at $J'=0$, the kinetic energy has its minimum at $x=0.5$
because of the optimal carrier density (Fig.~\ref{fig:e_x}). For $J'>0$,
however, the minimum of the kinetic energy is moved to larger doping
$x=5/8=0.625$, because this electronic filling allows to realize the FM
phase, while the magnetic correlations favor instead the CE phase at
$x=0.5$ at the expense of the kinetic energy. This demonstrates that
these two energies are to some extent complementary and their
competition controls the magnetic order. With finite nearest neighbor
Coulomb repulsion $V=t$, the kinetic energy of the CE phase is partly
lost due to charge order. Concerning the total energy, which contains
additional contributions from $V$ and $E_z$ and is not shown in Fig.
\ref{fig:e_x}, we want to emphasize that it is convex over the entire
doping range and for all parameter sets. Although this result seems to
exclude phase separation,\cite{Dag03} the clusters used in the present
study are definitely too small to address this issue in a conclusive
way.

\section{Magnetic phases in bilayer manganites}
\label{sec:bila}

\subsection{Phase diagrams for undoped system }
\label{sec:bila_x0}

Bilayer manganites like La$_{2-2x}$Sr$_{1+2x}$Mn$_2$O$_7$ represent
an intermediate situation between the 2D monolayer systems and 3D
perovskite manganites, so it is interesting to ask to what extent the
qualitative trends reported above for the monolayers are modified by the
interlayer coupling. We shall provide some limited answers to this
question, as unfortunately the calculations could only be performed for
certain selected fillings of the smallest bilayer
$\sqrt{8}\times\sqrt{8}\times2$ cluster used in the numerical studies.

\begin{figure}[t!]
\includegraphics[width=8.2cm]{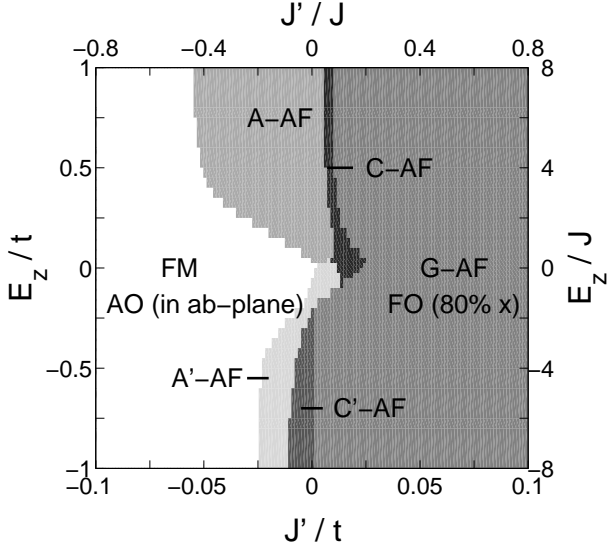}
\caption{
Phase diagram of the undoped bilayer system as obtained in $(E_z,J')$
plane with a $\sqrt{8}\times\sqrt{8}\times 2$ cluster at $T=0$.
The phases found between FM and $G$-AF order for two $ab$ planes with
interlayer coupling along $c$ axis are:
$C$-AF and $A$-AF phases as in Fig. \ref{fig:acg};
$C'$-AF:  FM along $c$ axis and AF within the $ab$ planes;
$A'$-AF:  FM along $a$ and $c$ axes and AF along $b$ axis.
The OO which accompanies the FM and $G$-AF phase at $E_z=0$ is also
indicated, and shown in
Figs.~\ref{fig:oo_bilayer}(a)--\ref{fig:oo_bilayer}(c).
Units $J'/t$ and $E_z/t$ correspond to $J=0.125t$.
}
\label{fig:phdbila}
\end{figure}

The study of magnetic correlations in an undoped ($x=0$) system
performed on $\sqrt{8}\times\sqrt{8}\times2$ clusters led to the phase
diagram in the $(J',E_z)$ plane (see Fig.~\ref{fig:phdbila}). It was
obtained by comparing the energies of various possible magnetic states
at $T=0$. As in a monolayer, large positive $J'$ leads to the $G$-AF
phase (with AF correlations in all three spatial directions), and
strongly negative $J'$ induces FM phase.\cite{notefm} Apart from these
two phases, one finds here several different types of intermediate
magnetic order in the crossover regime from FM to $G$-AF phase, with FM
bonds along either two or only one cubic direction, see Fig.
\ref{fig:acg}. Two phases of the same type, either $A$-AF and $A'$-AF,
or $C$-AF and $C'$-AF, are distinguished by the directions of FM bonds.
When the FM correlations occur within the $ab$ layers, as found
experimentally,\cite{Lin00} they are called $A$-AF and $C$-AF phases,
respectively. These phases appear for the expected\cite{Koi01} positive
($E_z>0$) crystal field which favors $|z\rangle$ orbital occupancy,
while for $E_z<0$ interlayer FM correlations occur in $A'$-AF and
$C'$-AF phases.

For intermediate $J'$, one obtains the $A$-AF phase, reminiscent of
the ground state of undoped 3D perovskite manganite LaMnO$_3$. For
$E_z\gtrsim 0.2t$ this plane is particularly robust
which suggests that positive $E_z$ simulates here the effect of missing
planes along $c$ axis on the electron distribution. Large $|z\rangle$
amplitude is needed as (i) this type of order is supported by the AO
order in the $ab$ planes, and (ii) $|z\rangle$ orbitals are responsible
for the AF interlayer coupling. This coexisting AO/FM phase within $ab$
planes is robust and can be suppressed only by AF core spin
superexchange $J'>0$ of a similar value as in the monolayer (see Fig.
\ref{fig:phd1}).

\begin{figure}[t!]
\includegraphics[width=8.0cm]{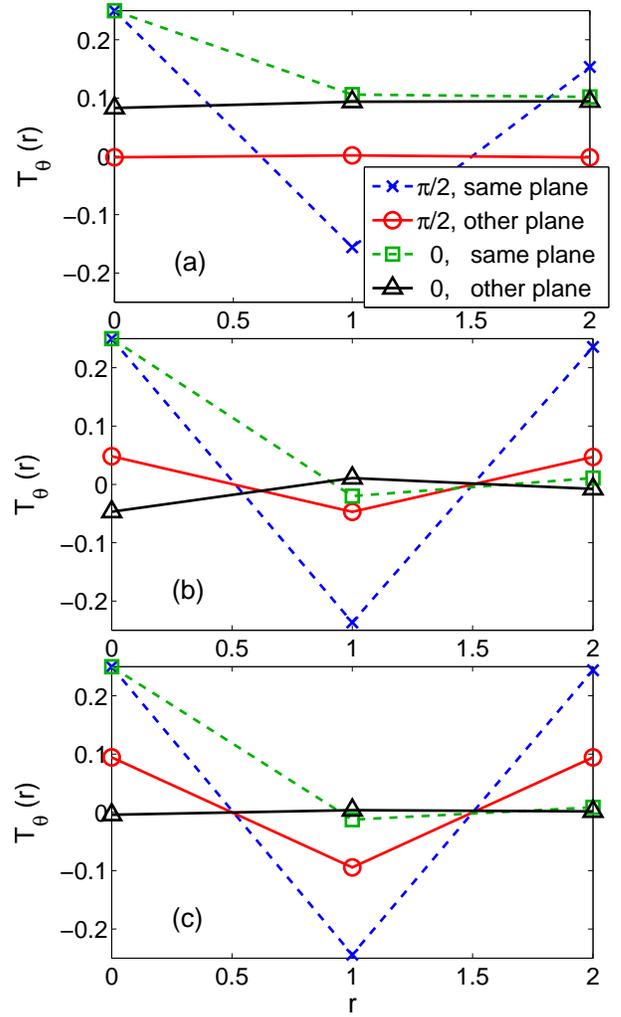}
\caption{(Color online)
Orbital correlations ${\cal T}_{\theta}(\vec r)$ (\ref{tco}) in $(01)$
direction within one $ab$ plane and between the two different planes of
the bilayer, as obtained for the undoped
$\sqrt{8}\times\sqrt{8}\times 2$ clusters at $E_z=0$ and $T=0$ in:
(a) $G$-AF phase,
(b) FM phase, and
(c) $A$-AF phase.
In each case $T_i^z$ operators are defined by different bases of
orthogonal orbitals: $\theta=0$ --- $\{|x\rangle,|z\rangle\}$, and
$\theta=\pi/2$ --- $\{|x\rangle+|z\rangle, |x\rangle-|z\rangle\}$.
Parameters: (a) $J'=0.05t$; (b) and (c) $J'=-0.02t$.
}
\label{fig:oo_bilayer}
\end{figure}

MC simulations performed at $\beta t=100$ for various parameter sets
support the phases found in the ground-state calculations ($J=t/8$ in
all cases):
$J'=0.05t$, $E_z=0$ ($G$-AF);
$J'=-0.02t$, $E_z=0$ (FM);
$J'=0$ and $J'=-0.02t$, $E_z=0.5t$ ($A$-AF);
$J'=-0.005t$, $E_z=-0.5t$ ($C'$-AF).
For some parameter sets close to the ground-state phase boundaries, we
observe competing phases, i.e., configurations showing several different
phases occur in the MC runs: $J'=0$, $E_z=0$ (FM and $A$-AF),
$J'=0.01t$, $E_z=0$ ($A$-AF, some configurations with $A'$-AF and
$C$-AF).

The crossover region from the FM to the $G$-AF phase with increasing
$J'$ is characterized by the competition between nearly degenerate
magnetic phases. As for the monolayer phase diagram, the ground-state
calculations also yield small regions with the $C$-AF or $C'$-AF phases.
The $C'$-AF was indeed found in MC runs for $J'=-0.005t$, $E_z=-0.5t$,
$\beta t=100$, $J=1/8 t$, but the $C$-AF phase (having a very narrow
stability region in the ground-state phase diagram) only occasionally
surfaced in the MC runs for $E_z =0$ and $J' = 0.01 t$ (competing with
$A$-AF) and $J'=0.02 t$ (competing with $G$-AF). It also occasionally
occurs at $J'=0.015t$, but the magnetic structure there remains unclear,
which could mean that temperature was still too high and/or the cluster
too small.

For negative $E_z$ in-plane $|x\rangle$ orbitals are favored, which in
turn enhances the region of stability of the $G$-AF phase. This can be
understood by looking at the orbital correlations in the $G$-AF phase
at $E_z=0$ depicted in Fig.~\ref{fig:oo_bilayer}(a): Even at orbital
degeneracy (without a crystal field), one finds rather pronounced
polarization for $\theta=0$ --- 80\% of the electrons occupy $|x\rangle$
orbitals as then the superexchange energy is gained. The $G$-AF phase
can therefore take advantage of a negative $E_z$, as seen in the phase
diagram of Fig. \ref{fig:phdbila}.

Figure~\ref{fig:oo_bilayer}(b) shows the orbital correlations in
the FM phase, where strong AO order within the layers coexists with FM
order in $ab$ planes. As this AO order can best develop when occupations
of $|x\rangle$ and $|z\rangle$ orbitals are nearly equal, the FM phase
is most pronounced around $E_z=0$. At the same time, the interlayer
coupling (kinetic energy) is much weaker. This state competes with the
$G$-AF state with larger occupancy of $|x\rangle$ orbitals.

\begin{figure}[t!]
\includegraphics[width=8.2cm]{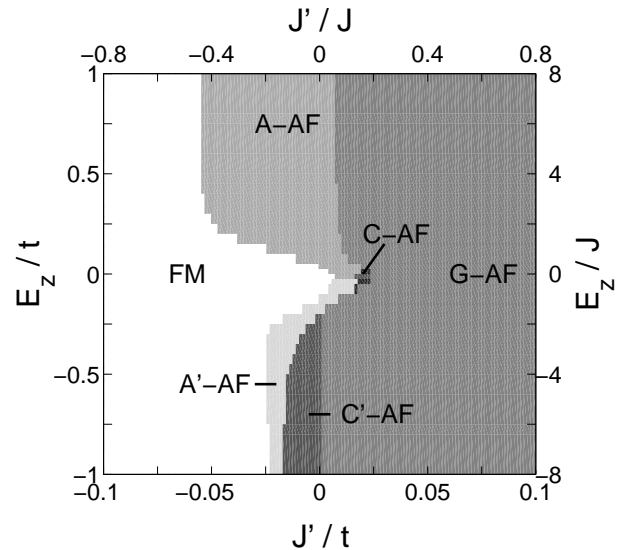}
\caption{
Phase diagram of the undoped $2\times 2\times 2$ cluster as obtained in
$(E_z,J')$ plane at $T=0$. The phases and the parameters are the same as
in the bilayer system (Fig. \ref{fig:phdbila}), with the directions of
FM bonds in $A$-AF, $A'$-AF, and $C'$-AF phases distinguished by the
symmetry-breaking crystal field $\propto E_z$, see Eq. (\ref{Hz}).
$C$-AF phase almost vanishes in this case.
}
\label{fig:phd222}
\end{figure}

Finally, the $A$-AF phase is found mainly for $E_z>0$ which enhances
electron density in $|z\rangle$ orbitals and supports the AF interlayer
coupling. The orbital correlations in this phase (at $E_z=0$) are
depicted in Fig. \ref{fig:oo_bilayer}(c) --- one finds the AO order
within the FM planes, i.e., orbital correlations for $\theta=\pi/4$ are
strongly alternating within the $ab$ planes. Between the planes, i.e.,
along the AF bonds in $c$-direction, the orbital correlation is weakly
positive, indicating weak FO order.

Although the phase diagram of Fig. \ref{fig:phdbila} was obtained with
small $\sqrt{8}\times\sqrt{8}\times 2$ clusters, we argue that it is
representative of the bilayer system in the thermodynamic limit, as it
is determined by short-range spin and orbital correlations that follow
from local superexchange interactions (charge excitations) on the
nearest-neighbor bonds. To support this point of view we present also
the phase diagram found with a smaller $2\times 2\times 2$ cluster in
Fig. \ref{fig:phd222}. One finds that indeed the same phases occur as in
Fig. \ref{fig:phdbila}, and their stability regimes are remarkably close
to those of the larger $\sqrt{8}\times\sqrt{8}\times 2$ cluster.

\subsection{Competition between different phases in half-doped
            bilayer clusters}
\label{sec:bilaha}

\begin{figure}[t!]
\includegraphics[width=8.0cm]{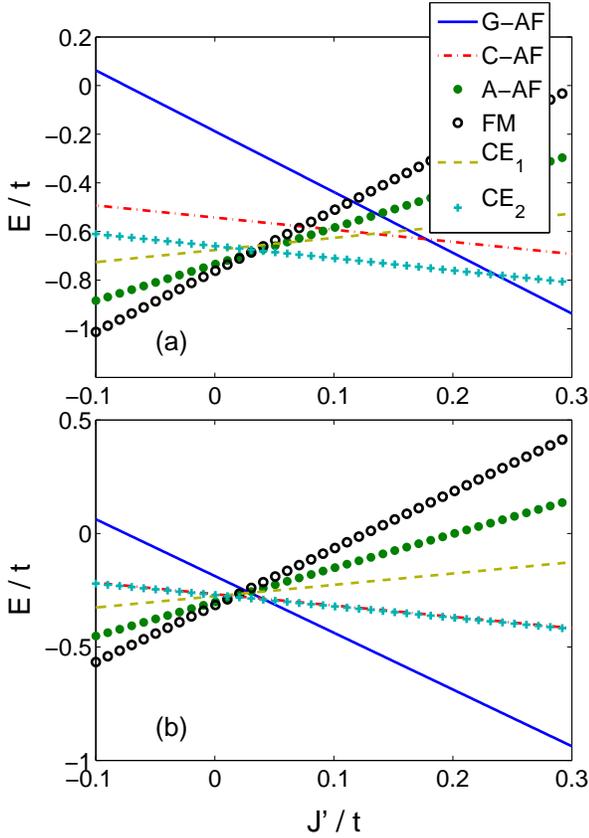}
\caption{(Color online)
Ground-state energy $E$ of various magnetic phases for increasing
$t_{2g}$ superexchange $J'$ in the half-doped bilayer
$\sqrt{8}\times\sqrt{8}\times 2$ cluster, as obtained at $T=0$ for:
(a) $V=0$, and
(b) $V=t$.
Different phases with coexisting AF and FM bonds are defined as follows:
$C$-AF --- FM in $a$ direction;
$A$-AF --- AF in $c$ directions, FM in $ab$ planes;
CE$_1$ --- FM zig-zag chains in $ab$ planes with FM interlayer coupling;
CE$_2$ --- FM zig-zag chains in $ab$ planes with AF interlayer coupling.
Parameters: $J=0.125t$ and $E_z=0$.
}
\label{fig:e05}
\end{figure}

Next, we investigate the magnetic and orbital order at half doping,
assuming orbital degeneracy ($E_z=0$). Figure
\ref{fig:e05}(a) shows the energy of various phases of the
half-doped bilayer depending on the value of the $t_{2g}$ superexchange
$J'$ in the absence of intersite Coulomb repulsion ($V=0$). Not only for
negative but also for small positive $J'$, the system is FM, while for
larger positive $J'>0.02t$ one finds $G$-AF phase. In between these
phases the CE phase, with alternating FM zig-zag chains in $ab$ planes
and AF coupling between them, has a lower energy. However, finite-size
effects are again important, as discussed in Sec.~\ref{sec:CE-mono}, and
thus one expects that instead the $C$-AF phase is the actual ground
state in the thermodynamic limit for the present realistic parameters.
(Similar to the 1D chains used for finite-size considerations instead of
2D clusters, one can use 2D ladderlike clusters instead of 3D clusters,
but as the attainable lengths are here much shorter definite results are
difficult to obtain. However, energy comparison of the $C$-AF and CE
phases on $4\times 2$ and $8\times 2$, including $6\times 2$ for $C$-AF
phase, suggests that the $C$-AF phase has lower energy.)

We have already shown in the case of a monolayer that the nearest
neighbor Coulomb repulsion $V$ decreases the range of stability of the
CE phase. Also for the bilayer system the CE phase is suppressed by
$V\sim t$, with weak preference of the $C$-AF phase [Fig.
\ref{fig:e05}(b)]. Namely, one finds that the energy of the
$C$-AF phase is lower than that of the CE phase for $V=t$ and
$J'>0.02t$. The reason is that the interlayer charge stacking (along $c$
direction) required by the CE phase costs extra energy now, whereas the
$C$-AF phase permits alternating charge order in all three directions
and has thus a lower energy-increment due to $V$. Unlike for the 2D
clusters of Sec. \ref{sec:phd}, the situation here is also similar for
$J=0.25t$ (not shown). We have verified that both phases here have very
similar energies on an $8\times2$ ladder, which seems to indicate that
the $C$-type phase wins in the thermodynamic limit for $J=t/4$ as well.
For $V=t$ and $J'<0.02t$ the $A$-AF phase is more stable than the CE
phase, which suggests that this parameter range is relevant for the
experimentally measured LaSr$_2$Mn$_2$O$_7$ sample.\cite{Med99,Lin00}

However, isotropic magnetic phases, FM for $J'<0.02t$ and $G$-AF for
$J'>0.02t$, are more stable than anisotropic phases in the entire range
of $J'$, if $V=t$ [Fig. \ref{fig:e05}(b)]. As for
monolayers, finite $V>0$ favors the $G$-AF phase, because its kinetic
energy vanishes already for $V=0$ and thus the energy of this phase is
not affected by $V$, while charge order induced by $V$ in other phases
hinders electron motion, and thus their total energies increase. While
the spin-orbital model leads to the CE phase in a monolayer at
$J\gtrsim t/8$ or for $V>0$, lattice degrees of freedom are apparently
needed to stabilize it in 3D perovskites. Furthermore, it should be
noted that the bilayer compound LaSr$_{2}$Mn$_2$O$_7$ shows $A$-AF
order,\cite{Med99,Lin00} in contrast to the monolayers and the 3D
compounds. The origin of this behavior remains unclear at present,
especially as the orbital correlations are reported to be similar
in all cases.\cite{Lar05,Yam00}

\begin{figure}[t!]
\includegraphics[width=8.0cm]{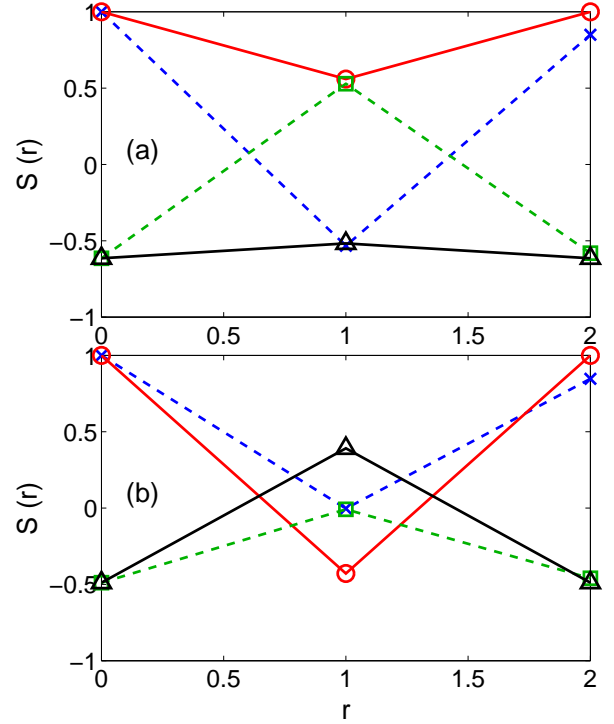}
\caption{(Color online)
Spin correlations (\ref{ss}) as obtained from MC simulations with a
$\sqrt{8}\times\sqrt{8}\times2$ cluster in the highly doped regime for
increasing coordinate $r$ in $ab$ plane:
(a) $G$-AF phase for one electron ($x=15/16$), and
(b) $C$-AF phase for three electrons.
Different symbols and lines indicate different types of neighbors:
$\times$ and dashed (blue) --- within the $ab$ planes along either
$(01)$ or $(10)$ direction;
$\circ$ and solid (red) --- within the $ab$ planes along the $(11)$
direction;
$\Box$ and dashed (green) --- between the two planes along $a$ or $b$
direction;
$\triangle$ and solid (black) --- between the two planes and along the
$(11)$ direction in $ab$ plane.
Parameters: $J=0.125t$, $J'=0.05t$, $V=E_z=0$, and $\beta t=100$.
}
\label{fig:ss_C}
\end{figure}

\subsection{Bilayer clusters at large doping}
\label{sec:bilala}

The reference system for large hole doping regime is $x=1$ case, when
itinerant $e_g$ electrons are absent and the magnetic order depends
exclusively on core spin superexchange $J'$ --- one finds then the
$G$-AF phase for $J'>0$ (while an unphysical $J'<0$ induces the FM
phase). We have been able to investigate the highly doped regime because the
small number of electrons leads here to a small Hilbert space which
allows to perform MC simulations down to $x\geq 3/4$ on a
$\sqrt{8}\times\sqrt{8}\times2$ cluster (filled by up to 4 electrons).
MC simulations show that $G$-AF order for $J'=0.05t$ is strong enough to
persist upon inclusion of one $e_g$ electron, i.e., at doping $x=15/16$,
as demonstrated by spin correlations presented in Fig.
\ref{fig:ss_C}(a). They describe $G$-type alternating spin order in
all three directions.

Somewhat lower doping $x=13/16$ (three electrons in the present cluster)
gives a $C$-AF phase with FM chains lying in the $ab$ planes, and
predominant occupation of directional orbitals along the FM direction
--- this state is stabilized by the double exchange mechanism. Indeed,
the strongly negative spin correlation at $(1,1)$ point, see Fig.
\ref{fig:ss_C}(b), is a signature of the $C$-AF phase. For distance
$r=1$ along $(0,1)$ direction, the signal is approximately zero, because
the FM and the AF correlations in the two directions nearly cancel each
other. Finally, the intermediate case of $x=0.875$ (with two electrons)
does not allow to establish a clear picture of magnetic correlations
(not shown), which may be due to strong competition between the two AF
phases. In fact, experiments show a transition from $G$-AF to $C$-AF at
doping $x\approx 0.9$ (see Ref. \onlinecite{Lin00}) which agrees with
these results.

\section{Discussion and conclusions}

The present study clarifies that orbital degrees of freedom are of
crucial importance for the understanding of magnetic correlations in
layered manganites. We treated a realistic model including intra- and
interorbital Coulomb interactions and investigated charge, intersite
spin and intersite orbital correlations in monolayer and in bilayer
manganites. The obtained results revealed a close relationship between
orbital and magnetic order which follows the Goodenough-Kanamori rules
at $x=0$.\cite{Goode} The magnetic phases found in different doping
regimes, where double exchange also contributes, are in accordance with
experiments over the whole doping range $0\leq x\leq 1$, particularly
for the monolayer systems.

For the undoped monolayers the model predicts either FM or AF order,
but not the $E$-AF phase, reported previously in an approach similar to
ours but ignoring on-site Coulomb repulsion between the $e_g$ electrons.
\cite{Dag03,Hot03} In their study it was stabilized by the kinetic
energy and arose mainly for \emph{nearly vanishing\/} electron-phonon
coupling,\cite{Hot03} i.e., in the situation when lattice degrees of
freedom could be neglected. Otherwise, the model of Ref.
\onlinecite{Hot03} is similar to our Hamiltonian apart from missing
Coulomb repulsion. Experimentally, however, the $E$-AF phase is only
observed\cite{Kim03} for the very \emph{strongly\/} JT distorted
HoMnO$_3$, and never in less distorted compounds. In the FM phase,
the OO induced by the orbital superexchange, i.e., by local Coulomb
repulsion, is found even in the absence of electron-phonon coupling, in
contrast to the model without Coulomb repulsion.\cite{Hot03} This shows
that the correct treatment of electron correlation effects due to large
Coulomb repulsion, which suppresses the kinetic energy in undoped
compounds (at $x=0$), is crucial for the qualitatively correct
description in this doping regime.

The experimental situation in doped La$_{1-x}$Sr$_{1+x}$MnO$_4$ could
be modeled with varying crystal field favoring out-of-plane $|z\rangle$
orbitals in the undoped system, and gradually decreasing with $x$ to
accelerate the electron transfer from $|z\rangle$ to $|x\rangle$
orbitals. Indeed, for positive $E_z\sim 0.5t$
the undoped monolayers contain then almost only $|z\rangle$ electrons,
while $|x\rangle$ occupation grows rapidly with doping when $E_z$
decreases in the present model, see Fig.~\ref{fig:nmc} in Sec.
\ref{sec:incr_doping}. Indeed, such a doping dependence of $E_z$ is
suggested by recent experiments.\cite{Sen05}

Another success of the model is that one observes the CE phase at half
doping with physically realistic parameters for layered manganites,
i.e., it is obtained for small $t_{2g}$ superexchange
$J'\gtrsim 0.03 t$, as deduced\cite{Ole05} from the analysis of
exchange constants in LaMnO$_3$. We also investigated the impact of
nearest neighbor Coulomb repulsion and found it to \emph{slightly}
stabilize the CE phase with respect to the $C$-AF phase in monolayers in
the relevant regime of $J'$, but to favor instead the $C$-AF phase in
bilayer clusters. (In both cases, the stability region of either the CE
or the $C$-AF phase shrinks and that of the $G$-AF phase grows when
nearest neighbor Coulomb repulsion is included.) In the CE
phase, we found relatively similar electron
densities at corner ($n^c\simeq 0.413$) and bridge ($n^b\simeq 0.587$)
positions in the zig-zag FM chains, which clearly contradicts the
localized picture of this phase. Furthermore, we observed that
electrons at the bridge sites are found in the directional
$3x^2-r^2/3y^2-r^2$ orbitals without crystal field and the planar
$z^2-x^2/y^2-z^2$ orbitals for $E_z>0$. 
While this is in contrast to the interpretation of some
experiments,\cite{Hua04,Wil05} other groups reported similar charge
distribution.\cite{Mar05} It is quite remarkable, however, that the CE
phase could be here explained by a purely electronic mechanism --- one
may expect that the oxygen distortions due to the JT effect would
further stabilize it. Also at large doping $x\simeq 0.75$, the
calculations for monolayer clusters predict the $C$-AF phase with
predominant occupation of directional orbitals, in agreement with
experimental data for Nd$_{1-x}$Sr$_{1+x}$MnO$_4$.\cite{Kim02}

An interesting variation of spin and orbital correlations with doping
was also found in the bilayer systems. They can be considered as
intermediate between 2D and 3D manganites, and we obtained the $A$-AF
phase for the realistic parameters at $x=0$, observed in undoped 3D
LaMnO$_3$ perovskite compound. The absence of the CE phase in the
bilayer phase diagram of Ling {\it et al.\/}\cite{Lin00} could not be
explained, however. Perhaps the electron transfer from $|z\rangle$ to
$|x\rangle$ orbitals at increasing doping is really fast, as suggested
by the variation of intralayer and interlayer exchange constants,
\cite{Per01,Ole03} and then the $A$-AF phase is stabilized again,
yet by different physical mechanisms. For very large doping $x>0.75$,
however, we obtained the $C$-AF and $G$-AF phases with a transition
between them, as indeed observed in bilayer compounds.\cite{Lin00}

Summarizing, the present study shows that the internal frustration of
magnetic interactions in doped manganites, with competing FM/AF terms
in the spin superexchange which coexist with complementary terms in the
orbital superexchange, has important consequences. Due to the intricate
energy balance between different types of intersite correlations, the
magnetic order may be completely switched over by small changes of
microscopic parameters, when the orbital order which coexists with it
switches at the same time. We find it quite encouraging that these
generic features, as well as the experimentally observed trends in
layered manganites, could be reproduced within the present microscopic
model.

\begin{acknowledgments}
We thank C. Baumann, B. B\"uchner, P. Horsch, and R. Klingeler for
insightful and stimulating discussions. This work has been supported
by the Austrian Science Fund (FWF), Project No.~P15834-PHY.
A.~M.~Ole\'s would like to acknowledge support by the Polish Ministry
of Science and Education under Project No.~1 P03B 068 26, and by COST Action
P16, ECOM. 
\end{acknowledgments}

\appendix*

\section{ slave boson approach for the 2D model of spinless fermions }

It is notoriously difficult to implement electron correlation effects
in nonmagnetic phases realized in multiband models. Therefore we
consider here a simpler case of a FM monolayer in the limit of large
$U\to\infty$ to compare the resulting charge distribution with the
exact diagonalization of finite 2D clusters. The electronic structure
for $e_g$ electrons is then described by the so-called orbital Hubbard
model of Ref. \onlinecite{Fei05}.

It was shown recently\cite{Fei05} that cubic invariance is obeyed when
the constraint of no double occupancy in the limit of large $U$ is
implemented by slave bosons for electronic states, using a basis of
complex orbitals at each site $i$,
\begin{equation}\textstyle{
|+\rangle_i=\frac{1}{\sqrt{2}}\big(|z\rangle_i - i |x\rangle_i\big),
\hspace{0.5cm}
|-\rangle_i=\frac{1}{\sqrt{2}}\big(|z\rangle_i + i |x\rangle_i\big).}
\label{complex}
\end{equation}
Then the hopping term $H_t^{\rm 2D}$ (\ref{eq:hopping}) and the
crystal-field term $H_z$ (\ref{Hz}) may be written as follows
(the superexchange terms vanish in the limit of $U\to\infty$):
\begin{eqnarray}
H_t^{U=\infty}&=& -\frac{1}{2} t
\sum_{\langle ij\rangle\parallel a,b}
  \Big[c_{i+}^{\dagger}c_{j+}^{}+c_{i-}^{\dagger}c_{j-}^{} \nonumber \\
&+&\Big(e^{-i\chi_{\alpha}}c_{i+}^{\dagger}c_{j-}^{}
       +e^{+i\chi_{\alpha}}c_{i-}^{\dagger}c_{j+}^{}\Big)
     +\mathrm{h.c.}\Big]                                   \nonumber \\
 &-&\frac{1}{2}E_z\sum_i(c_{i+}^{\dagger}c_{i-}^{}
                        +c_{i-}^{\dagger}c_{i+}^{}),
\label{Ht+-}
\end{eqnarray}
where $c_{i\pm}^{\dagger}$ are the corresponding creation operators,
and $\chi_a=+2\pi/3$, $\chi_b=-2\pi/3$ are the phase factors for the
bonds $\langle ij\rangle$ along $a$ and $b$ axis. Note that the crystal
field term contains only off-diagonal terms for the complex orbital
stares (\ref{complex}). In order to implement rigorously the constraint
at $U=\infty$ we replace now the fermion operators as follows,
\begin{equation}
c_{i\pm}^{\dagger} = b_{i\pm}^{\dagger}f_{i\mp}^{\dagger}e_i^{},
\label{krbosons}
\end{equation}
corresponding to a representation of the local states by
\begin{eqnarray}
|0\rangle_i&=&e_i^{\dagger}|{\rm vac}\rangle,   \nonumber \\
|+\rangle_i=c_{i+}^{\dagger}|0\rangle_i
          &=&b_{i+}^{\dagger}f_{i-}^{\dagger} |{\rm vac}\rangle,
          \nonumber \\
|-\rangle_i=c_{i-}^{\dagger}|0\rangle_i
          &=&b_{i-}^{\dagger}f_{i+}^{\dagger} |{\rm vac}\rangle,
\label{krstates}
\end{eqnarray}
where $|{\rm vac}\rangle$ is a true vacuum, following Ref.
\onlinecite{Fei05}. In the slave-boson mean-field approximation we
replace the boson operators by their averages, which leads to the
({\it a priori\/} site-dependent) hopping renormalization factors
$q_{i\pm}$. For an isotropic charge distribution one finds then
from a global constraint,
\begin{equation}
n_{+}+n_{-}=1-x,
\label{constraint}
\end{equation}
that the renormalization factors,
\begin{equation}
q_{i\pm}=\frac{x}{1-\langle f_{i\mp}^{\dagger}f_{i\mp}^{}\rangle}
   =\frac{x}{1-\langle n_{\pm} \rangle}=q_{\pm},
\label{gutzwiller}
\end{equation}
are the same for all kinetic energy terms. In this way one arrives at
the effective Hamiltonian with renormalized hopping terms,
\begin{eqnarray}
{\cal H}_{U=\infty}^{\rm MF}&=&
-\frac{1}{2}t\sum_{\langle ij\rangle\parallel a,b}
   \Big[q_+^{}\hat{f}_{i+}^{\dagger}\hat{f}_{j+}^{}
   +q_-^{}\hat{f}_{i-}^{\dagger}\hat{f}_{j-}^{}          \nonumber \\
&+&\sqrt{q_+q_-}\Big(e^{-i\chi_{\alpha}}
 \hat{f}_{i+}^{\dagger}\hat{f}_{j-}^{}
  +e^{+i\chi_{\alpha}}\hat{f}_{i-}^{\dagger}\hat{f}_{j+}^{}\Big)
  +\mathrm{h.c.}\Big]                                    \nonumber \\
&-&\frac{1}{2}E_z\sum_i(\hat{f}_{i+}^{\dagger}\hat{f}_{i-}^{}
                       +\hat{f}_{i-}^{\dagger}\hat{f}_{i+}^{}).
\label{effH}
\end{eqnarray}
By diagonalizing it in reciprocal space and using the inverse
transformation to Eq. (\ref{complex}),
\begin{equation}\textstyle{
|z\rangle_i=\frac{1}{\sqrt{2}}\big(|+\rangle_i + |-\rangle_i\big),
\hspace{0.5cm}
|x\rangle_i=\frac{i}{\sqrt{2}}\big(|+\rangle_i - |-\rangle_i\big).}
\label{complexinv}
\end{equation}
we determined the occupations of $|x\rangle$ and $|z\rangle$ orbitals
shown in Fig. \ref{fig:nmc}(b). A similar analysis used before for the
bilayer system gave the density distribution and the effective exchange
constants in good agreement with experiment.\cite{Ole03}



\begin{thebibliography}{00}
\bibitem{notefm} In reality, only $J'>0$ superexchange is allowed due
                 to the charge excitations by $t_{2g}$ electrons
                 between the Mn ions in high-spin states.
\bibitem{notece} As long as the FM chains in $ab$ planes are
                 geometrically oriented as shown in Fig.
                 \ref{fig:CE_orbs}(b).
\end{thebibliography}
\end{document}